\documentclass[11pt,preprint]{aastex}
\usepackage[varg]{txfonts}
\usepackage{graphicx, natbib, amssymb}
\usepackage{enumerate}
\bibpunct{(}{)}{;}{a}{}{,}
\usepackage[pagewise,displaymath, mathlines]{lineno}
\usepackage{pdflscape}

\newcommand{\kms}{km~s$^{-1}$}

\shorttitle{ISN~He: Mach numer, velocity vector, and temperature}
\shortauthors{Bzowski, Swaczyna et al. 2015}

\begin{document}

\title{Interstellar neutral helium in the heliosphere from IBEX observations.\\ III. Mach number of the flow, velocity vector, and temperature from the first six years of measurements}
    
\author{ M.~Bzowski\altaffilmark{1}, P.~Swaczyna\altaffilmark{1}, M.~A.~Kubiak\altaffilmark{1}, J.~M.~Sok{\'o}{\l}\altaffilmark{1}, S.~A.~Fuselier\altaffilmark{2,3}, A.~Galli\altaffilmark{4}, D.~Heirtzler\altaffilmark{5}, H.~Kucharek\altaffilmark{5}, T.~W.~Leonard\altaffilmark{5}, D.~J.~McComas\altaffilmark{2,3}, E.~M{\"o}bius\altaffilmark{5}, N.~A.~Schwadron\altaffilmark{5}, P.~Wurz\altaffilmark{4}}
\email{bzowski@cbk.waw.pl}

\altaffiltext{1}{Space Research Centre of the Polish Academy of Sciences, Warsaw, Poland}
\altaffiltext{2}{Southwest Research Institute, San Antonio, TX, USA}
\altaffiltext{3}{University of Texas at San Antonio, San Antonio, TX, USA}
\altaffiltext{4}{Physikalisches Institut, Universit{\"a}t Bern, Bern, Switzerland}
\altaffiltext{5}{Space Science Center and Department of Physics, University of New Hampshire, Durham, NH, USA}

\begin{abstract}
We analyzed observations of interstellar neutral helium (ISN~He) obtained from the Interstellar Boundary Explorer (IBEX) satellite during its first six years of operation. We used a refined version of the ISN~He simulation model, presented in the companion paper by \citet{sokol_etal:15b}, and a sophisticated data correlation and uncertainty system and parameter fitting method, described in the companion paper by \citet{swaczyna_etal:15a}. We analyzed the entire data set together and the yearly subsets, and found the temperature and velocity vector of ISN~He in front of the heliosphere. As seen in the previous studies, the allowable parameters are highly correlated and form a four-dimensional tube in the parameter space. The inflow longitudes obtained from the yearly data subsets show a spread of $\sim6\degr$, with the other parameters varying accordingly along the parameter tube, and the minimum $\chi^2$ value is larger than expected. We found, however, that the Mach number of the ISN~He flow shows very little scatter and is thus very tightly constrained. It is in excellent agreement with the original analysis of ISN~He observations from {\it IBEX} and recent reanalyses of observations from {\it Ulysses}. We identify a possible inaccuracy in the Warm Breeze parameters as the likely cause of the scatter in the ISN~He parameters obtained from the yearly subsets, and we suppose that another component may exist in the signal, or a process that is not accounted for in the current physical model of ISN~He in front of the heliosphere. From our analysis, the inflow velocity vector, temperature, and Mach number of the flow are equal to $\lambda_{\mathrm{ISNHe}} = 255.8\degr \pm 0.5\degr$, $\beta_{\mathrm{ISNHe}} = 5.16\degr \pm 0.10\degr$, $T_{\mathrm{ISNHe}} = 7440 \pm 260$~K, $v_{\mathrm{ISNHe}} = 25.8 \pm 0.4$~\kms, and $M_{\mathrm{ISNHe}} = 5.079 \pm 0.028$, with uncertainties strongly correlated along the parameter tube.
\end{abstract}


\section{Introduction}
\label{sec:Intro}
Analyses of interstellar neutral helium (ISN~He) measurements from the Interstellar Boundary Explorer (IBEX; \citet{mccomas_etal:09a, mobius_etal:09a, mobius_etal:09b}) by \citet{bzowski_etal:12a} and \citet{mobius_etal:12a} showed a very strong coupling of the He inflow velocity vector (speed and direction) and upstream temperature in a four-dimensionl (4D) ``tube'' of tightly coupled parameters \citep{mccomas_etal:12b}. These studies also produced the surprising conclusion that the statistically most likely solution for the flow of ISN~He observed during the 2009 and 2010 ISN observation seasons seemed to differ by $\sim4\degr$ in the inflow direction and $\sim3.5$~\kms~in speed from the solution obtained from a coordinated analysis of previous measurements \citep{mobius_etal:04a}, mostly based on direct sampling of ISN~He by the {\it GAS} instrument onboard {\it Ulysses} \citep{witte:04, witte_etal:04a}. For this different inflow vector, the consistent temperature along the {\it IBEX} tube was essentially identical to the GAS value of 6300~K. The same analysis by \citet{bzowski_etal:12a} also showed that the uncertainty range for the obtained solution was relatively wide along the parameter correlation tube in the four-dimensional parameter space, and that the direction and inflow velocity determined by \citet{witte_etal:04a} were not ruled out by the {\it {\it IBEX-Lo}} analysis, but statistically less likely. For the {\it Ulyses} flow vector, however, the temperature would have to be higher than the temperature found by \citet{witte:04} by $\sim2500$~K.

These findings by \citet{bzowski_etal:12a} and \citet{mobius_etal:12a} were of great impact and produced some controversy. If the inflow speed was really 3--4~\kms~slower than previously thought, then that would require a change in the configuration of the boundary region of the heliosphere compared with earlier views: the heliosphere would most likely feature no fast-mode bow shock \citep{mccomas_etal:12b, zank_etal:13a}, and so the disturbance of the local ISN gas by the heliosphere would spatially reach much farther upstream. In addition, \citet{frisch_etal:13a} considered the simplest explanation of the difference between the results of the measurements taken at different epochs, i.e., that the direction of the flow was changing with time. Their analysis of the published measurements and error bars carried out over the past 40 years suggested that the presence of a temporal change was statistically more likely than the lack of a change. In any case, \citet{bzowski_etal:12a}, \citet{mobius_etal:12a}, and \citet{mccomas_etal:12b} made it clear that the combined flow vector and temperature published for the Ulysses observations were inconsistent with the 4D tube of parameters required by the {\it IBEX} observations. This 4D tube remains intact today, as recently shown by \citet{mccomas_etal:15a} and \citet{leonard_etal:15a}, and as we show in this paper and other papers in this special issue \citep{mobius_etal:15b, schwadron_etal:15a}.    

Still, some researchers have voiced concerns about the early {\it IBEX} flow vector. \citet{lallement_bertaux:14a} suggested that the {\it IBEX} results were wrong because of errors in the data treatment by the {\it IBEX} team (specifically, that the effects of an instrument dead time were ignored) and possible inaccuracies in the modeling. Thus, they thought that the original ISN~He flow parameters and temperature from Ulysses were the correct ones and, consequently, considering any change in the inflow parameters with time was not needed. The misleading suggestions of a flawed data treatment by the {\it IBEX} team were rebutted by \citet{mobius_etal:15a}, \citet[][this volume]{mobius_etal:15b}, and \citet{frisch_etal:15a}. The mechanism of reduction in data throughput on board {\it IBEX} and the methods used to account for it in the data analysis are explained in detail by \citet[][this volume]{swaczyna_etal:15a}. 

Since the publication of the original analysis there has been an important development in our understanding of the {\it IBEX-Lo} measurements of ISN~He: \citet{kubiak_etal:14a} discovered the Warm Breeze -- a new population of neutral helium in the heliosphere. The Warm Breeze was shown to flow from ecliptic longitude between $230\degr$ and $250\degr$ and latitude between $8\degr$ and $18\degr$ at a speed between 7 and 15~\kms. Its temperature is between 7000 and 21\,000~K, with a most likely value of 15\,000~K, and the abundance relative to the primary ISN~He ranges from 4$\%$ to 10$\%$. The parent population assumed in the analysis was given by the homogeneous Maxwell-Boltzmann distribution function at 150~AU from the Sun. The two hypotheses for the origin of the Warm Breeze that have been identified as the most likely by the discoverers are as follow: (1) it is the secondary population of ISN He arising in the outer heliosheath due to the neutralization of interstellar He$^+$ ions via charge exchange with the pristine, undisturbed ISN He population flowing through the outer heliosheath, or (2) it is a non-thermalized flow of neutral He in the local interstellar matter that may originate at the hypothetic nearby boundary of the interstellar cloud which the Sun is embedded in and a medium with a much higher temperature, e.g., as in the hypothesis by \citet{grzedzielski_etal:10b} for the origin of the {\it IBEX} Ribbon. The discovery of the Warm Breeze is important in the context of finding the parameters of the primary ISN~He because, as we show in the paper, it modifies the ISN~He flux distribution observed by {\it IBEX} in all orbits. 

Another important development was the reanalysis of ISN~He observations by the GAS instrument on Ulysses. \citet{bzowski_etal:14a} carried out their analysis using the Warsaw Test Particle Model (WTPM), which also was used by \citet{bzowski_etal:12a} in their analysis of {\it IBEX} observations from 2009 and 2010. The analysis method was analogous to the method used in the {\it IBEX} analysis and the research team included the personnel who carried out the {\it IBEX} analysis and researchers who previously analyzed the {\it Ulysses} observations \citep{banaszkiewicz_etal:96a, witte_etal:96, witte_etal:04a, witte:04}. The data analysis included the entire set of data fom the {\it GAS} experiment, including the subset of data taken during the last orbit of Ulysses that ended in 2007, which had not previously been analyzed. The results indicated (to within errors) that the velocity vector of ISN~He from before does not feature any statistically significant changes with time, but that the temperature is much higher than had been obtained from the earlier analysis \citep{witte:04}. These findings were also confirmed by \citet{wood_etal:15a}, who used an analysis method for the Ulysses data analogous to the method used by \citet{lee_etal:12a} and \citet{mobius_etal:12a} to analyze the {\it IBEX} measurements.

Finally, \citet{leonard_etal:15a} analyzed measurements of ISN~He from 2012 to 2014, restricted to times when the spin axis tilt was less than $0.2\degr$ out of the ecliptic plane. These authors found that using data from times with large $(0.7\degr-4.9\degr)$ spin axis pointing out of the ecliptic gave quite different results, meaning that the analytic approximations used were not adequate for such pointing intervals. At the same time, \citet{mccomas_etal:15a} combined the \citet{leonard_etal:15a} analysis with Warsaw model calculations using data from 2013 and 2014. These authors found that the very tight coupling between the interstellar He inflow vector (longitude, latitude, speed) and the upstream temperature continued to define the same 4D tube in the parameter space as found earlier by \citet{mccomas_etal:12b}. They further indicated that the resolution of the apparent ``{\it Ulysses-IBEX} enigma'' was simply another location along the tube with a flow vector closer to the {\it Ulysses} value but a much higher temperature (by at least 1000~K). They suggested using combined {\it IBEX/Ulysses} values of $v_{\mathrm{ISN}} = \sim 26$~\kms, $\lambda_{\mathrm{ISN}} = \sim 255\degr$, $\beta_{\mathrm{ISN}} = \sim 5\degr$, and $T_{\mathrm{ISN}} = \sim 7000-9500$~K and discussed the implications of such a substantially warmer region of the interstellar medium than previously thought.

This paper is one of three papers presenting the latest elements of the analysis of ISN~He observed by {\it IBEX}, carried out using the analysis concept originally devised by \citet{bzowski_etal:12a}. The other two papers include a presentation of the WTPM model used to interpret the observations \citep{sokol_etal:15b} and of the uncertainty system for the data and parameter fitting method \citep{swaczyna_etal:15a}. The latter paper also presents corrections for the interface throughput reduction, described by \citet{mobius_etal:15a}, and a very accurately determined {\it IBEX} spin axis pointing for all analyzed orbits. These three papers are part of a larger and coordinated set of papers on interstellar neutrals as measured by IBEX, overviewed by \citet{mccomas_etal:15b} in this {\it Astrophysical Journal Supplement Series} Special Issue. 

In the present paper we analyzed data from all of the seasons of ISN~He observations from 2009 through 2014. For the first time, we included the influence of the Warm Breeze on the ISN~He flux in the analysis. Compared with the original analysis from 2012, we have improved and optimized WTPM, which allows us to analyze a data set more than four times larger than that in \citet{bzowski_etal:12a}. We have updated and refined the previously used model of ionization losses of ISN~He by \citet{bzowski_etal:13b} needed to correctly account for the losses of ISN~He inside the heliosphere during the observations covering a half of the solar cycle length. This update was presented by \citet{bochsler_etal:14a} and \citet{sokol_bzowski:14a}. We also reanalyzed the observational aspects of the measurements. In particular, we developed an analytical model of data throughput throttling effects due to the ambient electrons and refined the {\it IBEX} spin axis pointing determination, previously analyzed by \citet{hlond_etal:12a}. In addition, we collected all known uncertainties in the data and the correlations between them, and we construct a homogeneous, state-of-the-art system of accounting for the data uncertainties in the process of fitting the parameters of ISN~He gas. These aspects of the analysis are presented by \citet{swaczyna_etal:15a}. Since \citet{kubiak_etal:14a} pointed out the potential sensitivity of the simulated ISN~He signal to an energy sensitivity threshold, we also carried out an analysis of the energy sensitivity threshold of the {\it IBEX-Lo} instrument. We found that it is more than 20~eV and, consequently, does not affect the analysis of the measurements of ISN~He during the spring season of observations of ISN~He by {\it IBEX} \citep[][this volume]{galli_etal:15a, sokol_etal:15a}.

Our paper is complementary to the data analyses carried out using alternative methods, presented by \citet{leonard_etal:15a}, \citet{mobius_etal:15a}, and \citet{schwadron_etal:15a}. The analysis and results that we show here are an extension of the analysis presented by \citet{mccomas_etal:15a} based on the data collected in 2013 and 2014. 

\section{Data}
\label{sec:Data}
\begin{figure}
	\centering
	\includegraphics[scale=0.65]{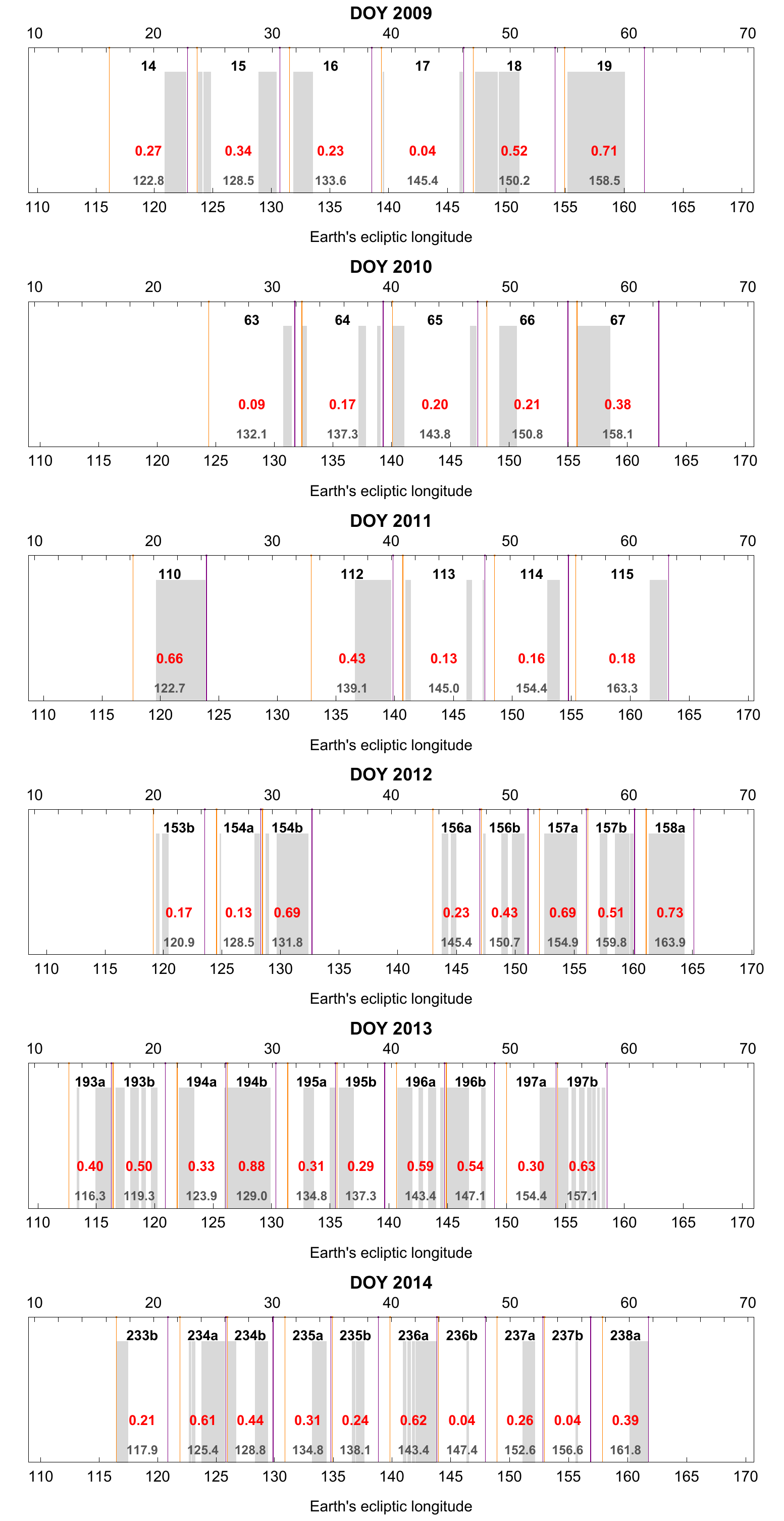}
	\caption{\tiny{Distribution of ISN good time intervals during the {\it IBEX-Lo} ISN~He observation seasons for the orbits taken into the analysis. The gray regions mark the individual good times intervals. The orange bars mark the beginning and the purple bars the end of the HASO intervals, when the measurements were actually carried out. The thick black labels mark individual orbits (or orbital arcs, in the 2012--2014 seasons). The numbers in the middle row mark the fraction  of HASO intervals occupied by the good time intervals for a given orbit. In the lower row, the approximate longitudes of the spacecraft during a given orbit are marked (actually, it is the Earth longitude averaged over the ISN good times for an orbit). They can be used to identify an approximate correspondence between orbits from different seasons. The upper horizontal axes are scaled in days since the beginning of the HASO interval for the first orbit for a given season, and the lower axis is ecliptic longitude.}}
	\label{fig:goodTimes}
	\end{figure}

In this study, we analyze data from {\it IBEX} energy step 2, histogram (HB) mode, taken during the ISN~He observation seasons 2009 through 2014. An individual data point is the count rate from an individual $6\degr$ bin in spin-angle, averaged over the ``good time'' intervals during an orbit, adopted from the ISN List data product. Details of the observation process relevant for our analysis are presented by \citet{swaczyna_etal:15a}. The initial data selection and cleaning, including identification of the good time intervals in the data, are presented by \citet{mobius_etal:12a} and \citet{leonard_etal:15a}. The ISN Good Times List is a data product maintained by the {\it IBEX} Science Team. It includes the time intervals when there were no issues with the synchronization of the spin pulse with the {\it IBEX} spin period, when Earth and the Moon were far from the field of view, when the count rate from magnetospheric sources outside the ISN signal angle range was low (assessing its magnitude inside the ISN range is challenging because the magnetospheric H is not distinguishable from the ISN atoms), and when the electron count rates, which are responsible for the interface throttling effect discussed by \citet{mobius_etal:15a} and \citet{swaczyna_etal:15a}, are not excessive. We also require a reliable spin axis pointing information from the {\it IBEX} navigation system. A graphical representation of the ISN good time interval distribution during the orbits used in the analysis is presented in Figure~\ref{fig:goodTimes}.

Each of the first six seasons of {\it IBEX-Lo} observations of ISN~He, each had some special features. 2009 was the first season and began directly after the commissioning of the spacecraft. Therefore, observations of the Warm Breeze, which typically start in the middle of November of the preceding year, could not be taken. In addition, the level of background due to electrons was elevated for the first few orbits (up to orbit 12), which was eliminated by a change of the on board processing program. 2010 was the first full observation season, the season from which \citet{kubiak_etal:14a} took the data to analyze the Warm Breeze. Just as in 2009, the observations were carried out using the regular mode of {\it IBEX-Lo}, with all 8 {\it IBEX-Lo} energy steps being stepped through in sequence. Data from one orbit, namely, the pre-peak orbit 62, were lost due to an on board computer reset. Because {\it IBEX} was optimized for its primary measurements of Energetic Neutral Atoms (ENAs) for the outer heliosphere \citep{mccomas_etal:09a, mccomas_etal:09c}, and not for ISN viewing, the ISN List coverage of the orbits with the highest count-rate region in Earth's orbit was relatively low. This potentially results in a relatively high statistical weight for the orbits with a lower intensity of the ISN~He signal, where the time coverage was better, but which likely include large contributions from the Warm Breeze or ISN~H.  

In 2011 and 2012, the instrument was operated in special modes during orbits 110 through 114, and 150a through 156a, respectively, and sensor-accumulated rates that are used in the throughput correction algorithm are not available at the required cadence. As a result, the ISN~He count rates from the ISN seasons 2011 and 2012 cannot be precisely corrected for the interface throughput reduction. However, as shown by \citet{swaczyna_etal:15a}, the ISN~He parameter determination is not seriously affected by this. Data from orbits 111 and 155 were lost because of on-orbit spacecraft issues. Additionally, the {\it IBEX} orbit was changed in the middle of 2011 to a more stable one against perturbations from the Moon \citep{mccomas_etal:11a}, and with a longer, $\sim9$-day period. Therefore, the {\it IBEX} spin axis repointing scheme also had to be modified, and after the orbit change the spin axis has been adjusted twice per orbit, close to the perigee and close to the apogee, and consequently the data from individual orbits are split into orbital arcs a and b. In addition, the target for the spin axis latitude was also modified. During the seasons 2009 through 2011, the spin axis was maintained $\sim0.7\degr$ above the ecliptic, and in 2012 it was alternated between $\sim -0.1\degr$ and $\sim0.7\degr$ \citep{hlond_etal:12a, swaczyna_etal:15a}.

In seasons 2013 and 2014, no data were lost and thus the coverage is approximately symmetric in ecliptic longitude around the maximum of the ISN~He flux. An important change, however, was a reduction in the post-acceleration voltage of the instrument because of instrumental reasons, which resulted in a reduction of the counting efficiency, and thus a reduction in the total ISN~He count rate. In 2013, the spin axis latitude was maintained between $\sim -0.2\degr$ and $\sim 0.4\degr$, and in 2014 it was alternated between $\sim 0\degr$ and $-4.9\degr$ in an effort to further strengthen the ISN observations. The good time coverage during the 2013 season is the best obtained so far.

Hence, measurements from different seasons are not precisely comparable to each other. Each season has its own specifics, which must be appropriately taken into account in the analysis. In particular, the equivalence of orbits between the seasons is limited because of the different good times distribution during the orbits and the non-identical distribution of the High altitude Science Operations (HASO) intervals from season to season. Also, the altitude at which HASO starts and ends, as well as the maximum distance of the spacecraft from the Earth differs, especially for the orbits from 2009 to 2011 compared to those from 2012 to 2014. 

An important aspect of data selection is maintaining the homogeneity of the sampled population. \citet{mobius_etal:12a} and \citet{bzowski_etal:12a} noticed the presence of the Warm Breeze during the early orbits of the season and realized that the ISN~H signal is expected at later times during each season (end of March/beginning of April). As a consequence, they decided to focus their analysis of ISN~He on those orbits for which the contributions from these populations are expected to be minimal. After the analysis of the Warm Breeze by \citet{kubiak_etal:14a} and the ISN~H by \citet{saul_etal:12a} and \citet{saul_etal:13a}, and in particular by \citet{schwadron_etal:13a}, it became clear that observations taken from ecliptic longitudes larger than $\sim160\degr$ include a component of ISN~H (see also, e.g., Figure~3 in \citet{kubiak_etal:13a}). \citet{leonard_etal:15a} analyzed measurements of ISN~He from 2012 to 2014 restricted to orbits when the spin axis tilt out of the ecliptic was less than $0.2\degr$ and suggested that an important Warm Breeze contribution was present in the data from all orbits with ecliptic longitude less than $\sim115\degr$ (see also \citet[][this volume]{mobius_etal:15b}. They recommended analyzing the ISN~He signal based on a subset of orbits restricted to the interval of ecliptic longitudes of spin axis $115\degr - 160\degr$, i.e., approximately from the second week of February until mid-March. We follow this recommendation, which is supported by our own analysis which follows. 

{\it IBEX} spins around its own axis, which is maintained close to the Sun and near the ecliptic plane. The {\it IBEX-Lo} boresight scans a great circle of $360\degr$ in the sky, and spin-angles from $180\degr$ to $360\degr$ point to the ram hemisphere, i.e., to the hemisphere that contains the direction of the Earth motion around the Sun. The peak of the primary ISN~He component enters the instrument around spin-angle $264\degr$. The signal due to a single Maxwell-Boltzmann population of ISN~He is expected to closely follow a Gaussian shape as a function of the spin-angle. Hence, it should be symmetric about the spin-angle corresponding to the peak flux. Any additional population should manifest itself by breaking this symmetry, unless it happened to be symmetric about identical spin-angles for all of the orbits in question, as in the case of the ISN population, which seems unlikely. Therefore, a good indicator for the presence of an additional population in the signal is a disturbed symmetry of the observed signal. 

To test the symmetry of the signal, we analyze its central moments. A moment $m^{(n)}$ of the $n$-th order (i.e., the $n$-th moment) is defined as
	\begin{equation}
	m^{(n)}=\sum_{i}{\psi_{i}^{n}F_{i}} / \sum_{i}{F_{i}}
	\label{eq:defMoment}
	\end{equation}
where $\psi_{i}$ is the $i$-th spin-angle during a given orbit and $F_{i}$ is the measured count rate for this bin.	
	
The first moment $m^{(1)}$ corresponds to the ``center of mass'' of the signal and for a symmetric distribution in the spin-angle in which it coincides with the peak value. The central moments $M^{(n)}$ of the distribution are defined as 
	\begin{equation}
	M^{(n)}=\sum_{i}{F_{i}\left( \psi_{i} - m^{(1)} \right)^{n}} / \sum_{i}{F_{i}}.
	\label{eq:defCentralMoment}
	\end{equation}
Using this definition, the first central moment is always 0. For a symmetric distribution, the square root of the second central moment is proportional to the width of the distribution, and the third central moment corresponds to the skewness of the distribution and is expected to be zero due to the symmetry. 

We found that a small positive skewness occurs in the simulations carried out under the assumption of a single homogeneous Maxwell-Boltzmann population in the gas ahead of the heliosphere. Generally, however, for Earth longitudes larger than $\sim 115\degr$ the skewness should disappear if only the primary He component is contributing to the observations. We took this as a guideline for the data selection: we calculated the central moments and their uncertainties from the data using the full uncertainty system developed by \citet{swaczyna_etal:15a}. We compared the skewness of the observed ISN~He count rates with the Warm Breeze subtracted and without it subtracted. The expectation was that the data with the Warm Breeze contribution not subtracted could show some skewness, but subtraction of the Warm Breeze should take it out.

\begin{figure}
	\centering
	\resizebox{\hsize}{!}{\includegraphics{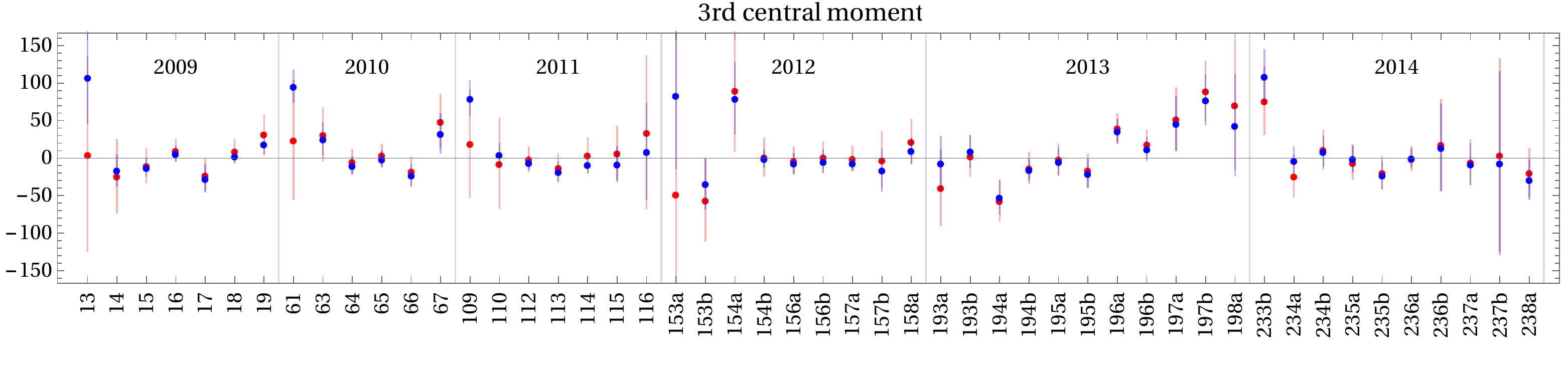}}
	\caption{Illustration of the third central moment of the data, shown from left to the right for the 2009 through 2014 ISN~He observation seasons. The vertical bars separate the seasons. The error bars were calculated using the full uncertainty system presented by \citet{swaczyna_etal:15a}. Red dots represent the skewness of the data with the contribution from the Warm Breeze subtracted from the observed signal ans blue dots the skewness of the data without the Warm Breeze subtraction. The horizontal axis represents the orbit numbers, the vertical axis is scaled in degree$^3$, as implied for the third central moment by the defining Equation~\ref{eq:defCentralMoment}.}
	\label{fig:skewness}
	\end{figure}

In reality, it turned out that some orbits at the beginning and the end of the ISN~He observation seasons show a residual skewness even after subtraction of the Warm Breeze. Aside from some exceptions, subtracting the Warm Breeze, calculated precisely for the observation conditions for a given season with the parameters found by \citet{kubiak_etal:14a}, gives a smaller skewness but does not eliminate it altogether within the error bars at the early and late orbits of the season, i.e., at the orbits for which the contribution from the Warm Breeze is expected to be the largest. This is illustrated in Figure~\ref{fig:skewness}, which shows the skewness of the data for all seasons with uncertainties calculated using the full uncertainty system from \citet{swaczyna_etal:15a}. Evidently, subtracting the Warm Breeze eliminates the skewness very well for season 2010: only orbit 67 departs from 0 within the uncertainty range after correction, whereas all orbits except 65 show a statistically significant skewness before correction. This is understandable, since \citet{kubiak_etal:14a} determined the Warm Breeze parameters based solely on the data from the 2010 season. Subtracting the Warm Breeze also reduces the skewness satisfactorily also for seasons 2009, 2011, and 2014. For the 2012 season, the remnant skewness is largest in the early and late orbits during the season. An exception is season 2013, for which subtracting the Warm Breeze alleviates the skewness only a little in the early orbits and in fact increases it in the late orbits. 
	
This analysis suggests that the choice of the range of spin axis ecliptic longitudes from $115\degr$ to $160\degr$ made by \citet{leonard_etal:15a} and \citet{mobius_etal:15b} is good for the studies of the pristine ISN~He. It spans $45\degr$, and to approximately match this range also in the other dimension, i.e., in ecliptic latitude, we use spin-angle bins from $252\degr$ to $282\degr$. In this way, we symmetrically straddle the peak of the count rate, which occurs between the bins corresponding to spin-angle $264\degr$ and $270\degr$, and we cut out the far-tail data points from all of the orbits, which have the largest remnant contribution from the Warm Breeze (see Figure~\ref{fig:WarmBreezeContri}) and ISN~H. We applied these criteria consistently to all six observation seasons. We verified that the data selected in all of the orbits adopted for the analysis fit a Gaussian shape (i.e., departures do not exceed $2 \sigma$). 
	\begin{figure}
	\centering
	\resizebox{\hsize}{!}{\includegraphics{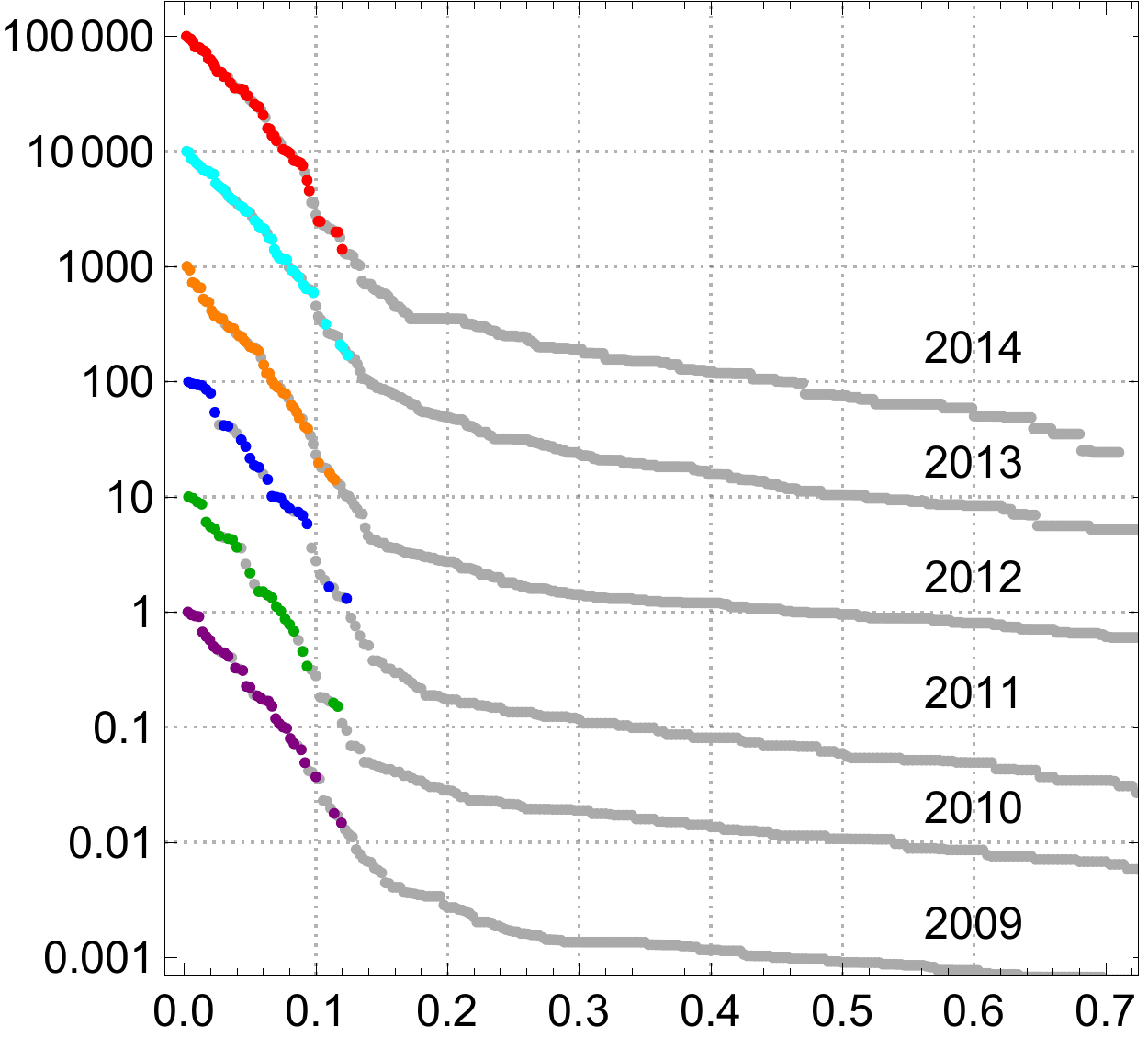}}
	\caption{Ratios of count rates in individual spin-angle bins to the maximum count rate registered during a given season, shown on a logarithmic scale. The gray points represent all of the data points (count rates in each $6\degr$ Histogram Bin, averaged over good time intervals) from the orbits adopted for analysis during a given season, and the colored points mark the points actually taken for analysis. The points are sorted for each season in decreasing order, and scaled to the maximum value registered during the season, so that the maximum value corresponds to 1 in the vertical scale. Seasons 2010 through 2014 are systematically spaced by a factor of 10 for clarity. The horizontal axis is linearly scaled so that the total number of points from the adopted orbits from a given season corresponds to 1.}
	\label{fig:dataOrdered}
	\end{figure}

The final data selection is presented in Figure~\ref{fig:dataOrdered}, which corresponds to Figure~1 in \citet{sokol_etal:15a}. This figure illustrates the contrast of the data points adopted for ISN~He analysis relative to the background. The count rates are shown season by season, sorted in decreasing order, and scaled to the maximum value registered during the given season (so that the season maximum corresponds to 1). Data from individual seasons are presented vertically spaced a decade apart, to facilitate viewing. The data range corresponding to the ISN signal is clearly visible as the sharply decreasing sequence of points on the left-hand side of the figure. The background corresponds to the data portion starting from approximately 0.25 on the horizontal scale. The region between $\sim0.15$ and $\sim0.25$ on the horizontal scale corresponds to a mixture of ISN~He and the Warm Breeze, in varying proportions. Note that as a result of the adopted data selection, we leave out the data points with count rates less than approximately $2\%$ of the seasonal maximum, and that the percentage of points used in the analysis is approximately equal from season to season.

The data used in this study, along with their uncertainties and the ancillary information needed to reproduce the results, are part of the {\it IBEX} Data Release~9. Details are provided by \citet{schwadron_etal:15a}, \citet{swaczyna_etal:15a}, and \citet{sokol_etal:15b}. 

\section{Model fitting}
\label{sec:ModelFit}
As the physical model, we adopted the distribution function of ISN~He ahead of the heliosphere in the form of a Maxwell-Boltzmann function in the reference frame comoving with the unperturbed flow of interstellar gas, with the velocity vector and temperature homogeneous in space. We assume that these conditions apply at a finite distance from the Sun, which we adopted at 150~AU. The rationale for this is explained in \citet{sokol_etal:15b} (see also further discussion in \citet{mccomas_etal:15b}). This assumption is identical to that adopted by \citet{bzowski_etal:12a}. The atoms from the population enter the heliosphere and follow the Keplerian hyperbolic trajectories due to the gravitational attraction from the Sun and suffer ionization losses due to their interaction with the solar EUV radiation and solar wind. Details of the calculation of the simulated signal are presented by \citet{sokol_etal:15b}. In order to include the Warm Breeze in the model, we adopt the model presented by \citet{kubiak_etal:14a} and calculate the count rates expected for a given data set. This simulated signal is then subtracted from the data. The Warm Breeze is taken into account throughout the entire analysis presented in this paper. The consequence of neglecting it in the analysis are discussed by \citet{swaczyna_etal:15a}, who identified it as the main contributor to the very high values of reduced $\chi^2$ obtained by \citet{bzowski_etal:12a} in their original analysis of the first two ISN~He seasons.

The baseline model fitting method is presented by \citet{swaczyna_etal:15a} in Section~4. In brief, it is a multi-tier minimization of the difference between the data and the model, scaled by the magnitude of the data uncertainty. This difference is denoted as $\chi^2$ and defined in Equation~16 in their paper. The optimum parameter set corresponds to the global minimum value of $\chi^2$. To find this optimum parameter set, i.e., the ecliptic longitude $\lambda$ and latitude $\beta$ of the inflow direction, the inflow speed $v$, and the ISN~He temperature $T$ ahead of the heliosphere, the model is first calculated on a carefully selected grid of 4297 points in the parameter space. The grid is deployed around a set of equally spaced ecliptic longitude values. For all of the ecliptic longitudes sampled, the grid points with the lowest $\chi^2$ values are identified and the respective minimum $\chi^2$ values are found for these longitudes by the minimization of a 3D paraboloid in the parameter space, fitted to a limited subset of points around the minima for each $\lambda$ value. In the last step, the global optimum parameter set is found from a $\chi^2$ minimization for the model searching for the minimum in $\chi^2$ as a function of $\lambda$. With this minimum identified, the final search for all 4 parameters is performed.  

We also used a supplementary method, which approximately corresponds to the method originally developed by \citet{lee_etal:12a} and applied by \citet{mobius_etal:12a}, \citet{leonard_etal:15a}, and \citet{mobius_etal:15b}. In this method, the measured signal is assumed to conform to a Gaussian function of the spin-angle, and the search of the ISN~He parameters is based on the analysis of the parameters of this Gaussian function, fitted to a time series of the measurements in each orbit. Thus, the symmetry of the signal around the peak is implicitly assumed and the analysis seems to be less prone to a bias from asymmetries due to the hypothetical presence of an unaccounted population in the signal. 

We have adopted a similar approach as an alternative to our direct-fitting method. For the selected orbits and selected spin-angle range $240\degr-294\degr$, we take the signal summed over all of the good time intervals for a given orbit and fit the Gaussian function of the spin-angle at each orbit in the following form:	
	\begin{equation}
	F\left( \psi \right) = b + f \exp \left[ -\left( \psi - \psi_0 \right)^2 / \sigma^2 \right]
	\label{eq:GaussFunDef}
	\end{equation}
where $b$ corresponds to the background, $f$ to the peak height of the signal, $\psi_0$ to the peak position in spin-angle, and $\sigma$ to the peak width. The Gaussian function parameter fitting is done by $\chi^2$ minimization using the uncertainty system described by \citet{swaczyna_etal:15a}. For each orbit, the Gaussian fit is described by four parameters, and thus the model has a number of parameters equal to four times the number of orbits. Thus, for each orbit $j$, we obtain a set of local ISN~He beam parameters $\{b_j, f_j, \psi_{0j}, \sigma_{j}\}$ in the {\it IBEX} reference frame. The covariance matrix form was obtained as described by \citet{swaczyna_etal:15a}, thus the correlations between bins and orbits are also included in the Gaussian representation of the signal. Subsequently, we perform global ISN~He parameter fitting by $\chi^2$ minimization, adopting as the data this series of Gaussian beam parameters, and as the model the series of the Gaussian parameters $\{f_j, \psi_{0j}, \sigma_{j}\}$ with their covariance matrix, obtained from fitting Equation~\ref{eq:GaussFunDef} to the simulated signal for a given ISN~He parameter set. The uncertainties are obtained in a fashion similar to the direct fit method.

\section{Results}
\label{sec:Results}
\begin{figure}
	\centering
	\resizebox{\hsize}{!}{\includegraphics{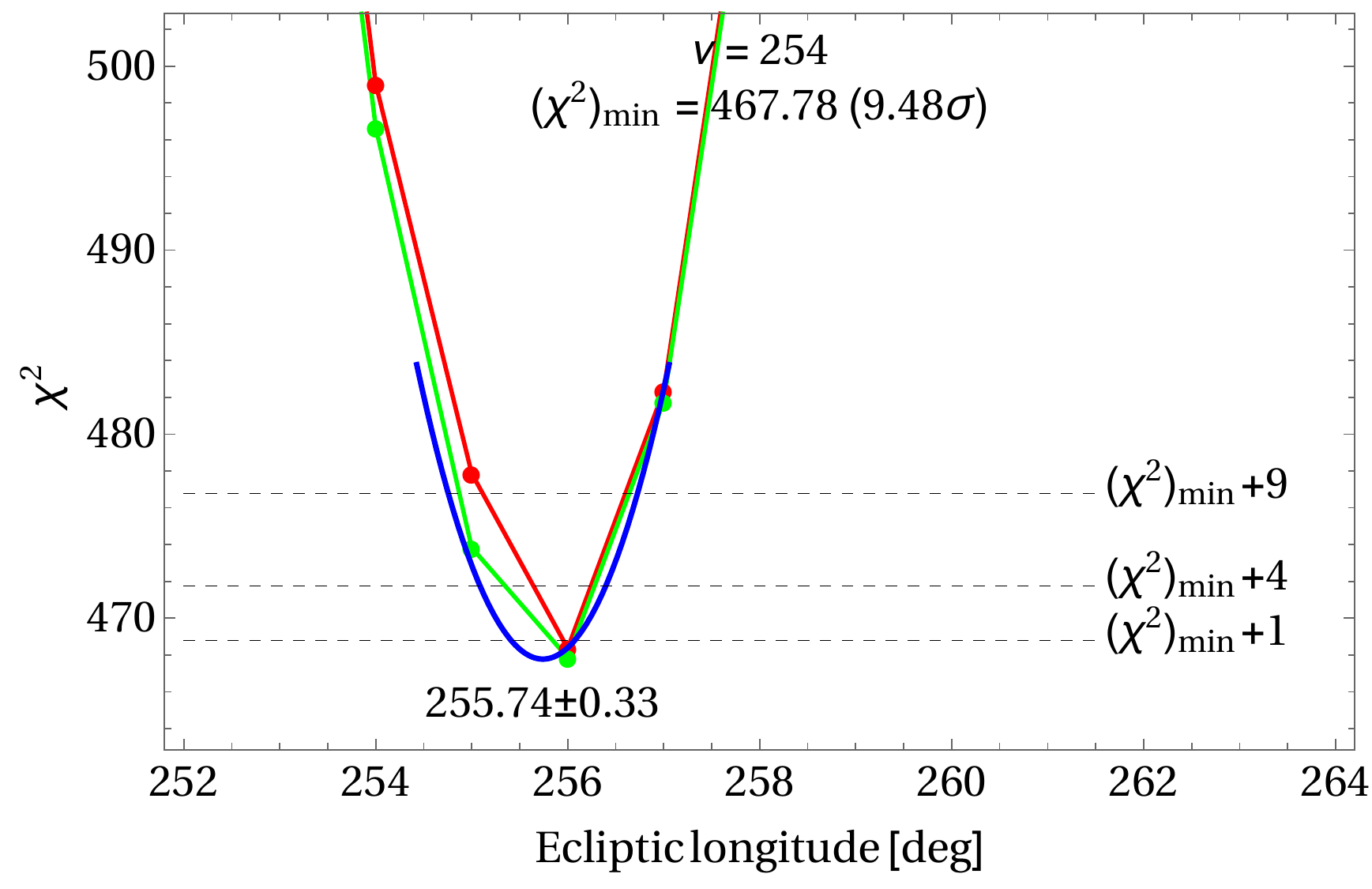}}
	\caption{Minimized $\chi^2$ as a function of the ecliptic longitude of the ISN~He inflow direction $\lambda$ (blue line). Red points mark $\chi^2$ minimized over the parameter grid points for a given $\lambda$. Green points are $\chi^2$ values for fixed longitudes, minimized with respect to the other three parameters. The blue line is a parabola corresponding to the center line of the covariance ellipsoid, defined in Equation~\ref{eq:allYearsDirCov}, straddling the optimum solution. The broken lines mark the levels equal to the minimum $\chi^2$ value plus 1, 4, and 9, respectively. See the text for an explanation of the other symbols. See also the caption for Figure~\ref{fig:allYearsDirCorLine}.}
	\label{fig:allYearsDirChi2}
	\end{figure}

We fit the ISN~He parameters to the data from all six observation seasons over the spin-angle range from $252\degr$ to $288\degr$, for all orbits in the Earth ecliptic longitude range $115\degr$ to $160\degr$, fitting the model directly to the observed count rate (with the contribution from the Warm Breeze subtracted). We found that the inflow velocity vector of ISN~He obtained from the fit is $\lambda_{\mathrm{ISNHe}} = 255.75\degr \pm 0.33\degr$, $\beta_{\mathrm{ISNHe}} = 5.16\degr \pm 0.07\degr$ in the J2000 ecliptic coordinates, $v_{\mathrm{ISNHe}} = 25.76 \pm 0.27$~\kms, and temperature $T_{\mathrm{ISNHe}} = 7440 \pm 190$~K. The inflow direction in galactic coordinates is $l_{\mathrm{ISNHe}} = 3.77\degr \pm0.16\degr$, $b_{\mathrm{ISNHe}} = 14.94\degr \pm 0.30\degr$, with the correlation coefficient between the uncertainties in the two coordinates equal to $-0.86$. A plot of $\chi^2\left(\lambda\right)$ for this calculation is presented in Figure~\ref{fig:allYearsDirChi2}. The correlation and covariance matrices for the best-fit parameters are given in Equations~\ref{eq:allYearsDirCorr} and \ref{eq:allYearsDirCov}:

	\begin{equation}
	\mathrm{\textbf{Cor}} = \left( 
	\begin{array}{c|cccc}
	 & \lambda & \beta & T & v \\ \hline
	 \lambda & 1.000 & -0.326 & -0.913 & -0.892 \\
	 \beta & -0.326 & 1.000 & 0.285 & 0.294 \\
	 T & -0.913 & 0.285 & 1.000 & 0.950 \\
	 v & -0.892 & 0.294 & 0.950 & 1.000 \\
	\end{array}
	\right)
	\label{eq:allYearsDirCorr}
	\end{equation}
	\begin{equation}
	\mathrm{\textbf{Cov}} = \mathbf{\Sigma}_{\mathrm{ISNHe}} = \left( 
	\begin{array}{c|cccc}
	 & \lambda~\left[\degr\right] & \beta~\left[\degr\right] & T~\left[10^3~\mathrm{K}\right]& v~\left[\mathrm{km~s}^{-1}\right]\\ \hline
	 \lambda~\left[\degr\right] & 0.10734 & -0.00787 & -0.05641 & -0.07965 \\
	 \beta~\left[\degr\right] & -0.00787 & 0.00544 & 0.00396 & 0.00591 \\
	 T~\left[10^3~\mathrm{K}\right] & -0.05641 & 0.00396 & 0.03553 & 0.04879 \\
	 v~\left[\mathrm{km~s}^{-1}\right] & -0.07965 & 0.00591 & 0.04879 & 0.07421 \\
	\end{array}
	\right).
	\label{eq:allYearsDirCov}
	\end{equation}

In addition to the ISN He parameters, we calculated the scaling factors to rescale the simulated flux expressed in the physical units to count rates. These parameters, calculated separately for each observation season, were being searched simultaneously with the ISN parameter fitting using the least-squares $\chi^2$ fitting procedure presented by \citet{sokol_etal:15b} in their Section 2.7. For the seasons 2009 through 2014, we obtained the scaling factors equal to $1.676\cdot 10^{-5}$, $1.666\cdot 10^{-5}$, $1.511\cdot 10^{-5}$, $1.450\cdot 10^{-5}$, $7.709\cdot 10^{-6}$, and $7.977\cdot 10^{-6}$, respectively. These quantities, albeit technical in nature, provide an important insight into the stability of the instrument sensitivity and the fidelity of the ionization loss model, as we detail in the discussion section.

\begin{figure}
	\centering
	\begin{tabular}{ccc}
	\includegraphics[scale=0.4]{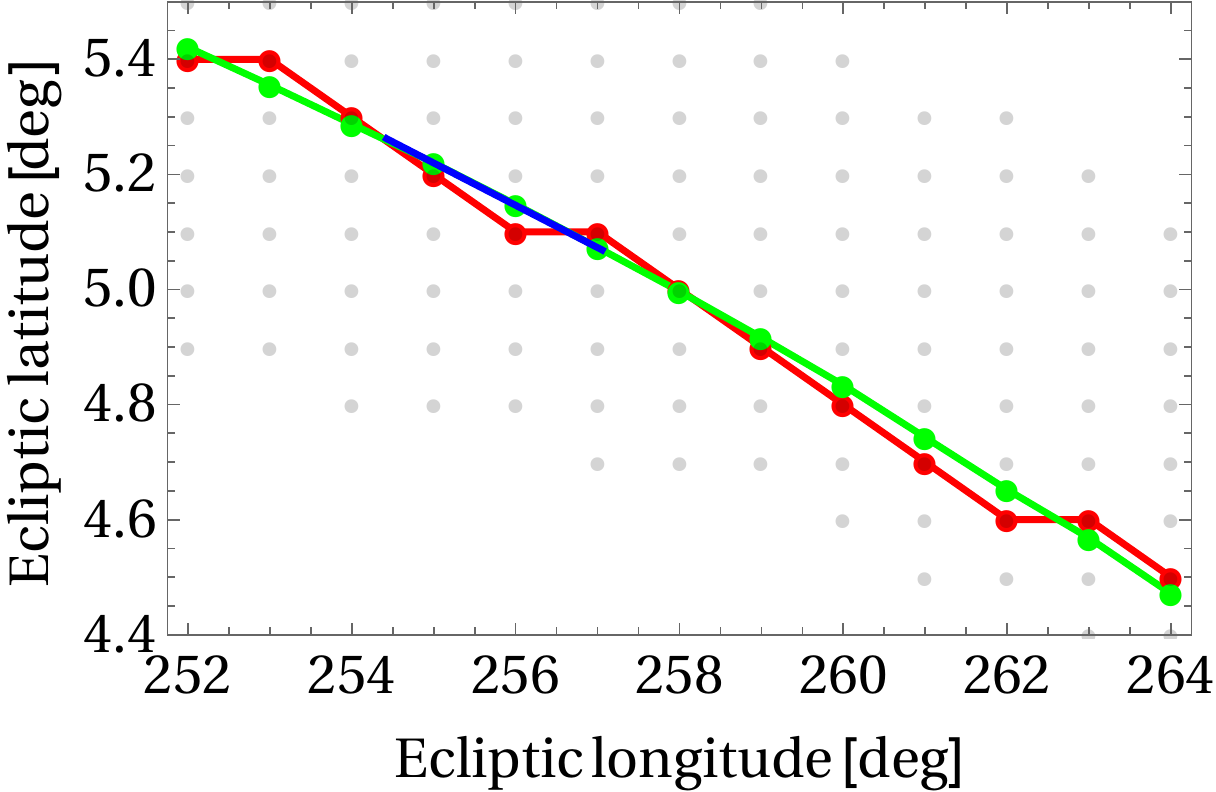} & \includegraphics[scale=0.4]{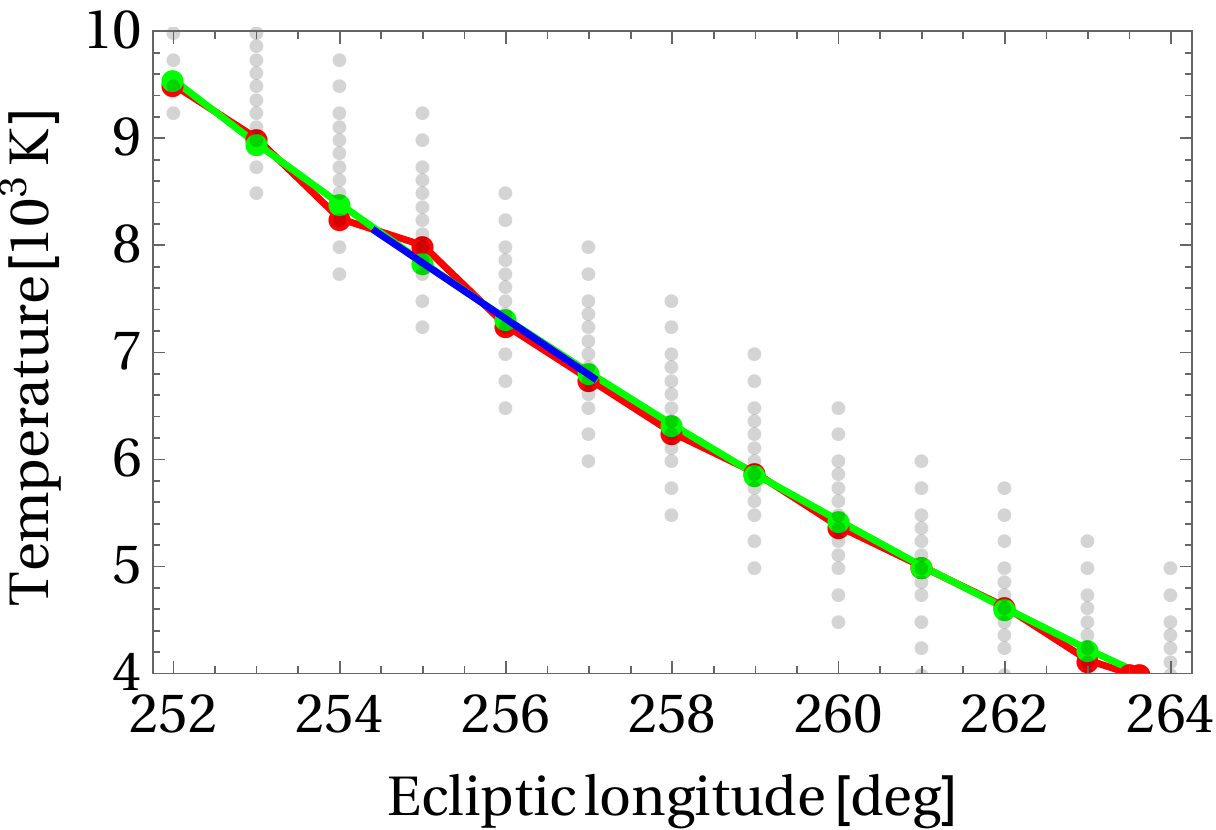} & \includegraphics[scale=0.4]{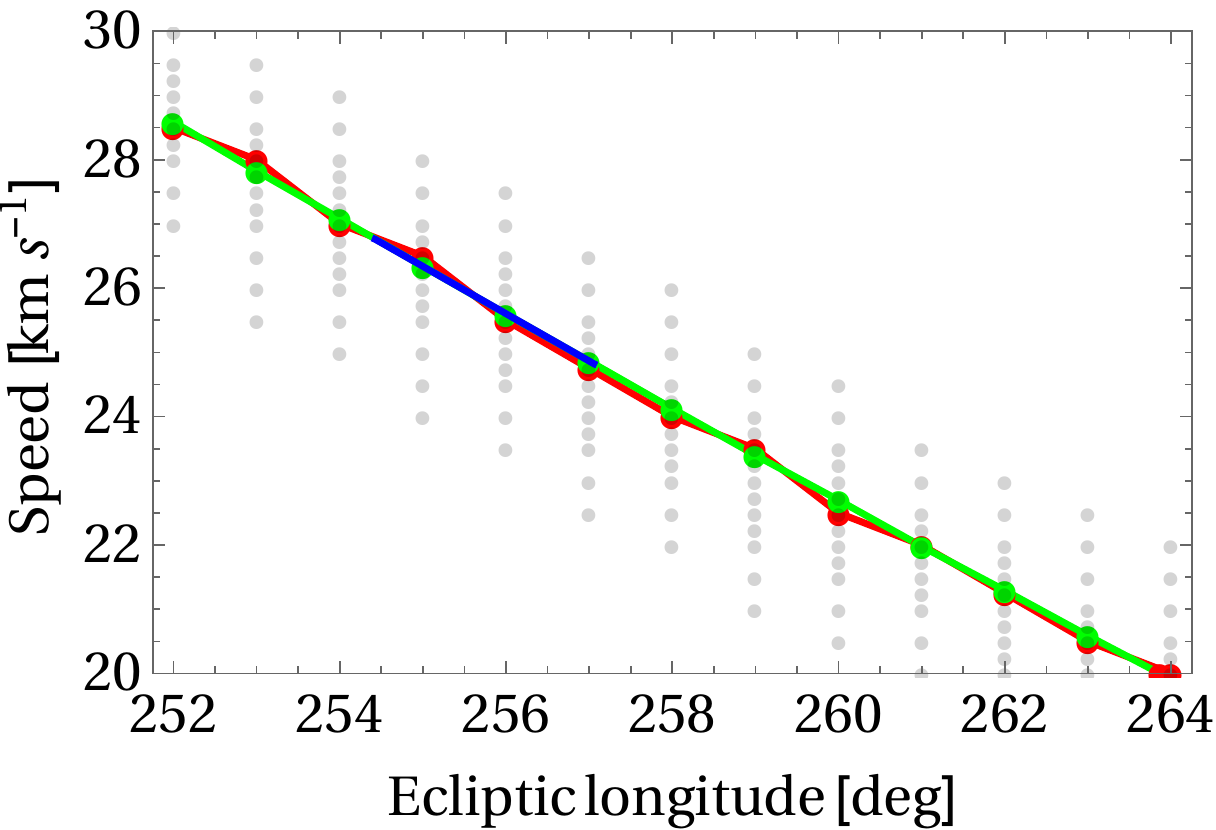}\\
	\end{tabular}
	\caption{Parameter correlation lines in three cuts through the four-dimensional parameter space show the 4D correlation tube. Shown as a function of inflow longitude are the lines of $\chi^2$ minimized over all parameters except for ecliptic latitude (left panel), temperature (middle panel), and speed (right panel). The gray dots represent the nodes of the entire computation grid and the red points those grid nodes for which the minimum value of $\chi^2$ was found for a given $\lambda$. The green points represent the results of inter-grid optimization for $\lambda$ values corresponding to the calculation grid and the other parameters marginalized. The green line, which is a projection of the green line from Equation~\ref{eq:allYearsDirCorr} on the respective parameter pair subspaces, is provided to guide the eye. The blue line represents the center line of the ellipsoid of covariance, defined in Equation~\ref{eq:allYearsDirCov}, which straddles the optimum solution and corresponds to the parabola drawn in Figure~\ref{fig:allYearsDirChi2}.}
	\label{fig:allYearsDirCorLine}
	\end{figure}

\begin{figure}
	\centering
	\begin{tabular}{cc}
	\includegraphics[scale=0.4]{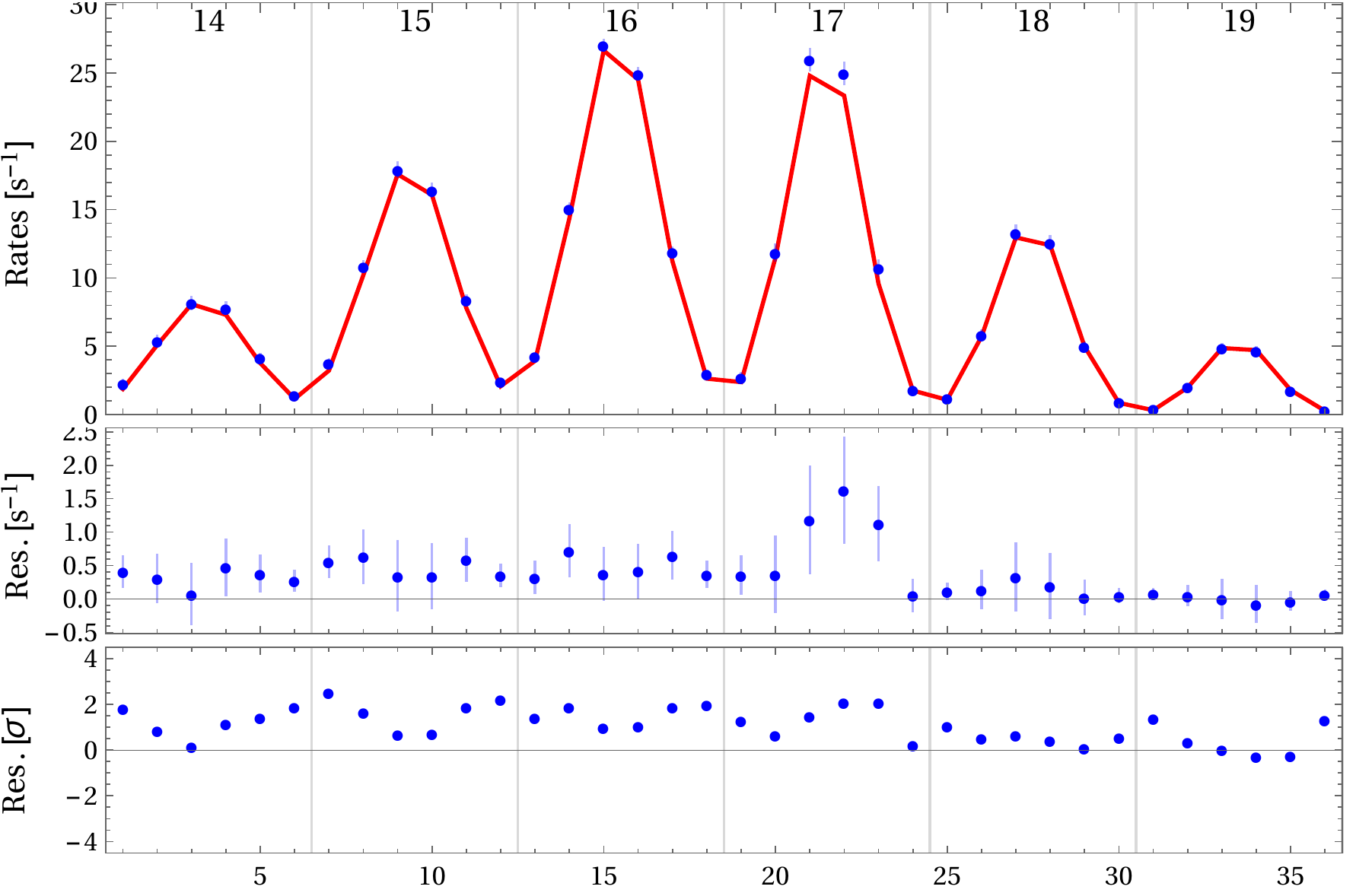} & \includegraphics[scale=0.4]{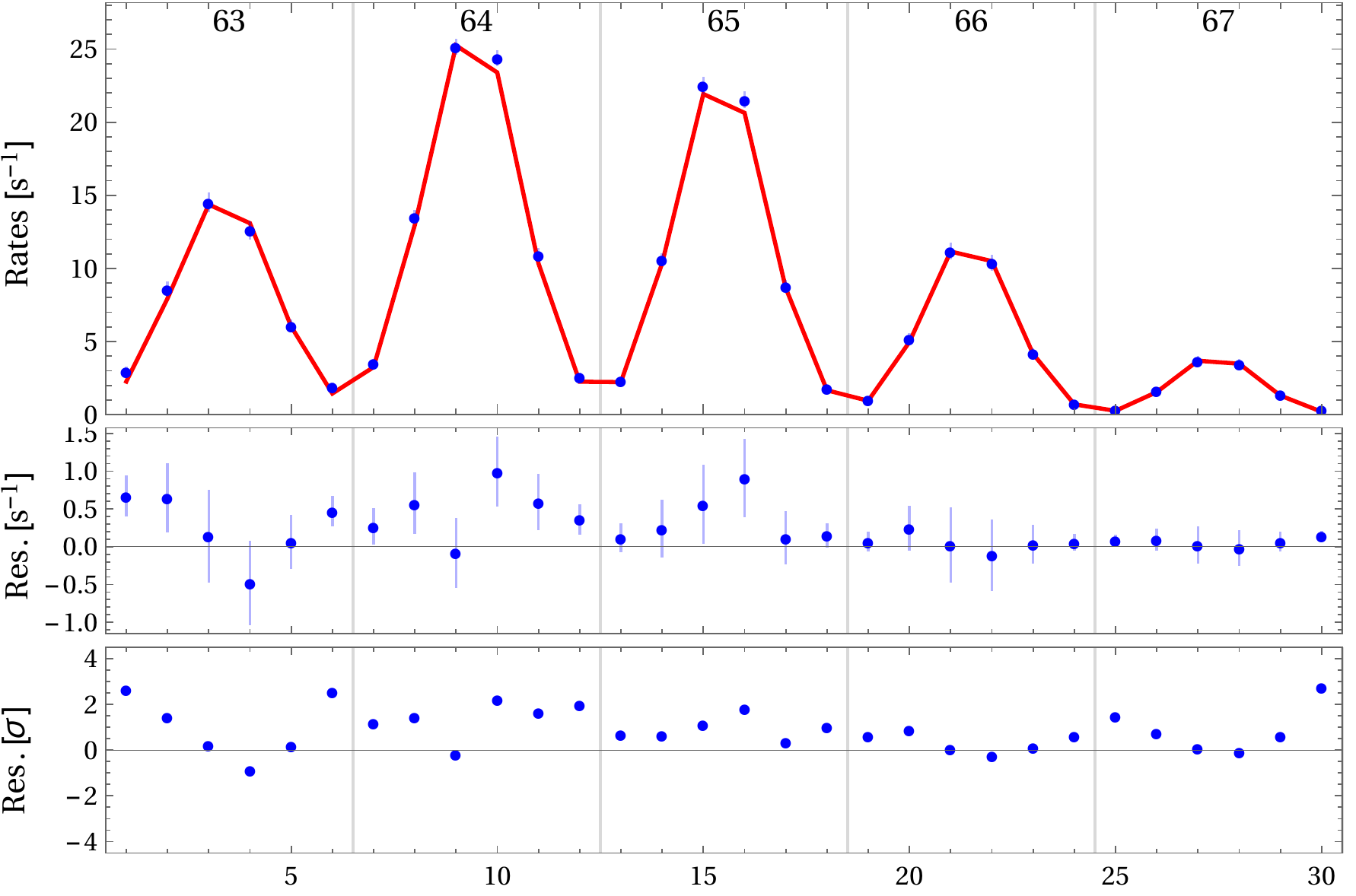}\\
	\includegraphics[scale=0.4]{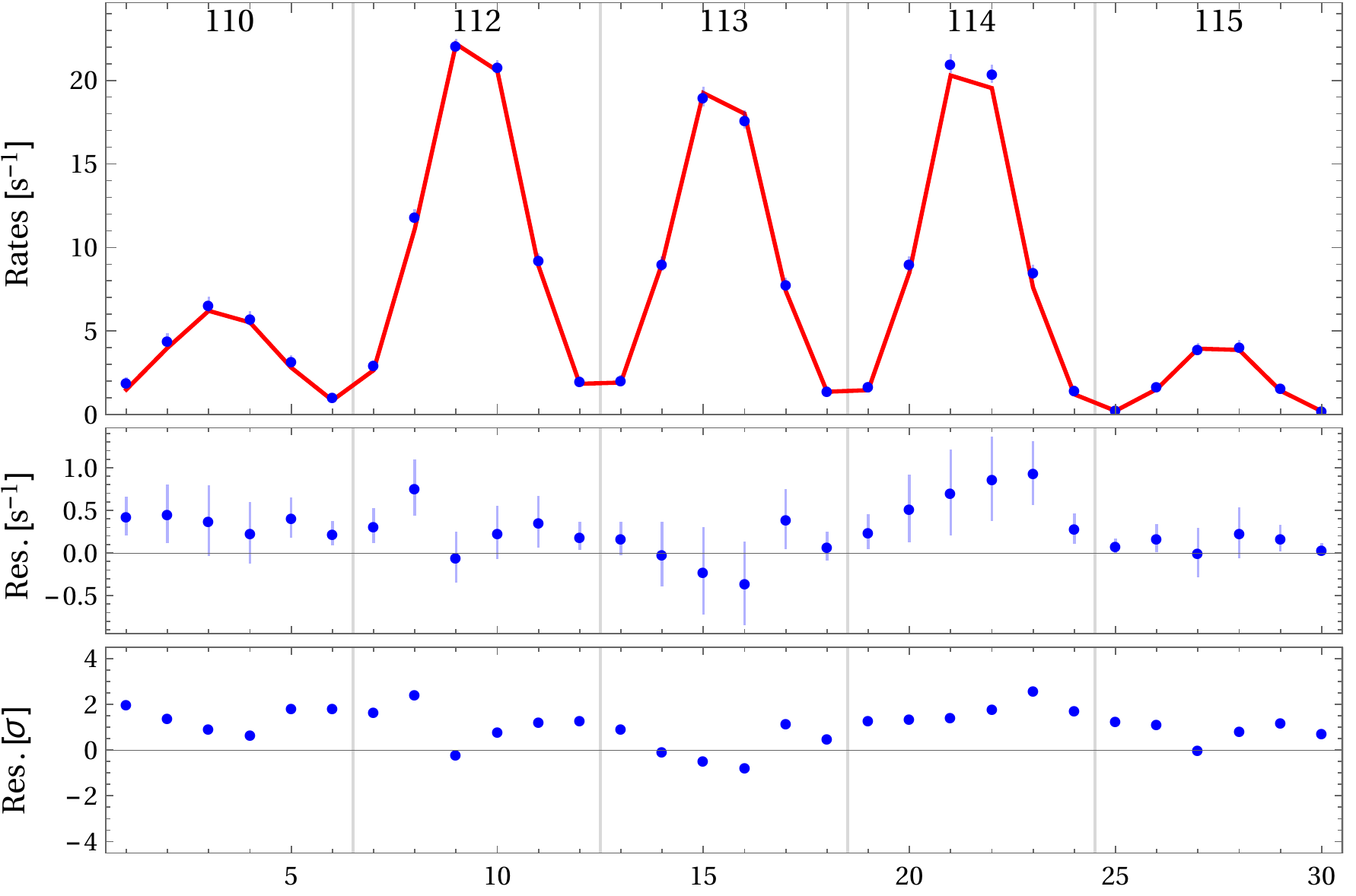} & \includegraphics[scale=0.4]{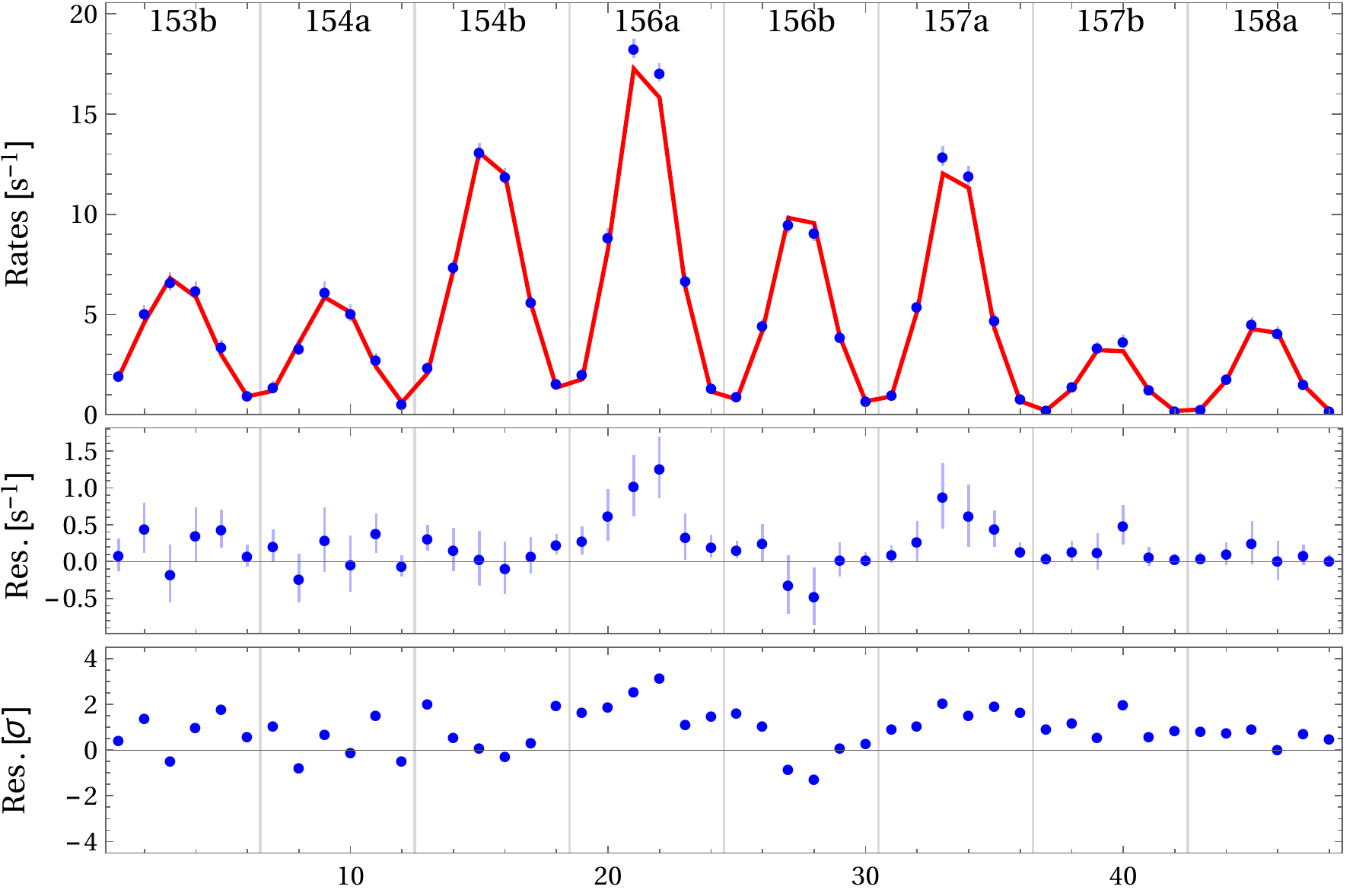}\\
	\includegraphics[scale=0.4]{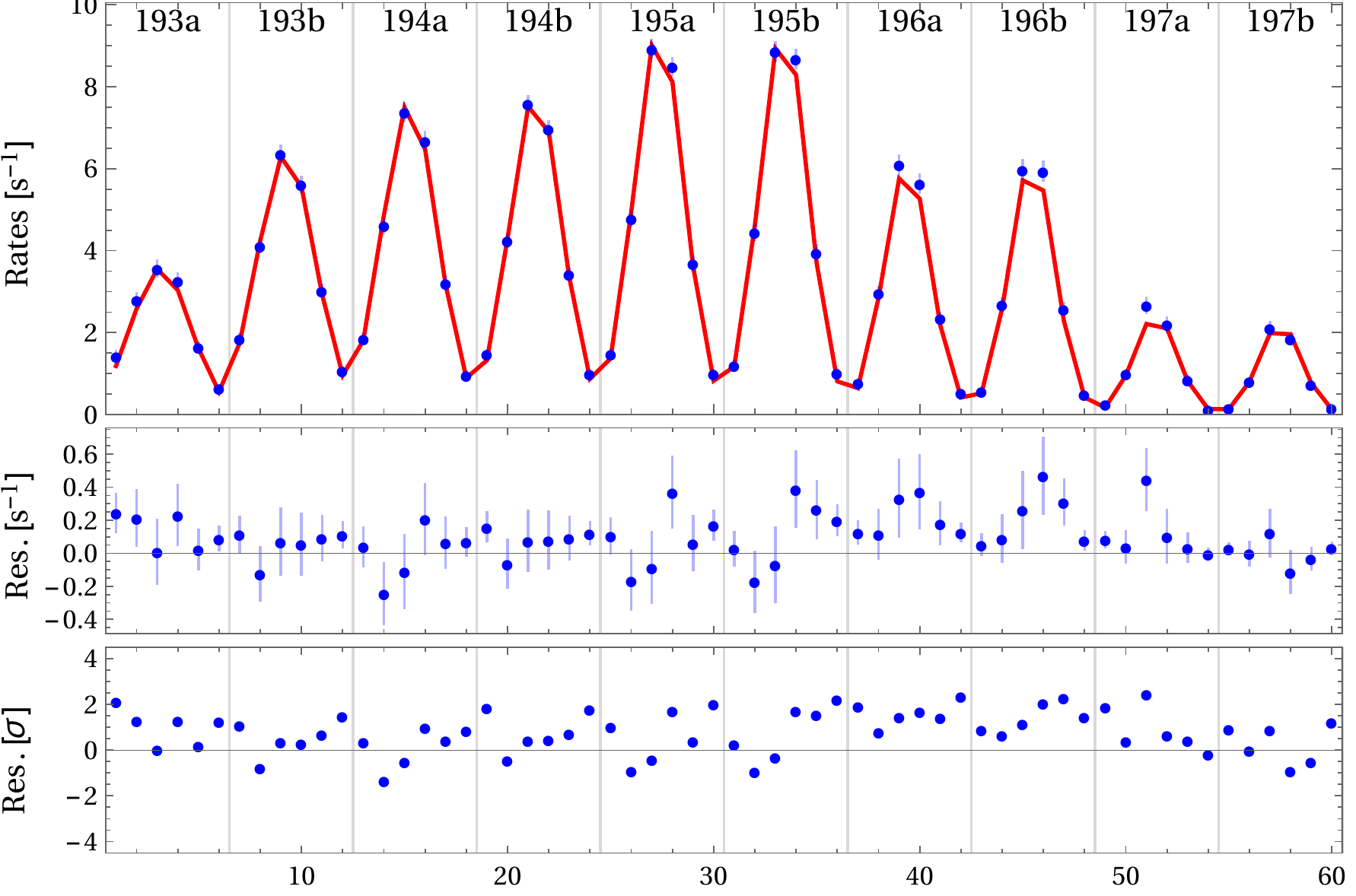} & \includegraphics[scale=0.4]{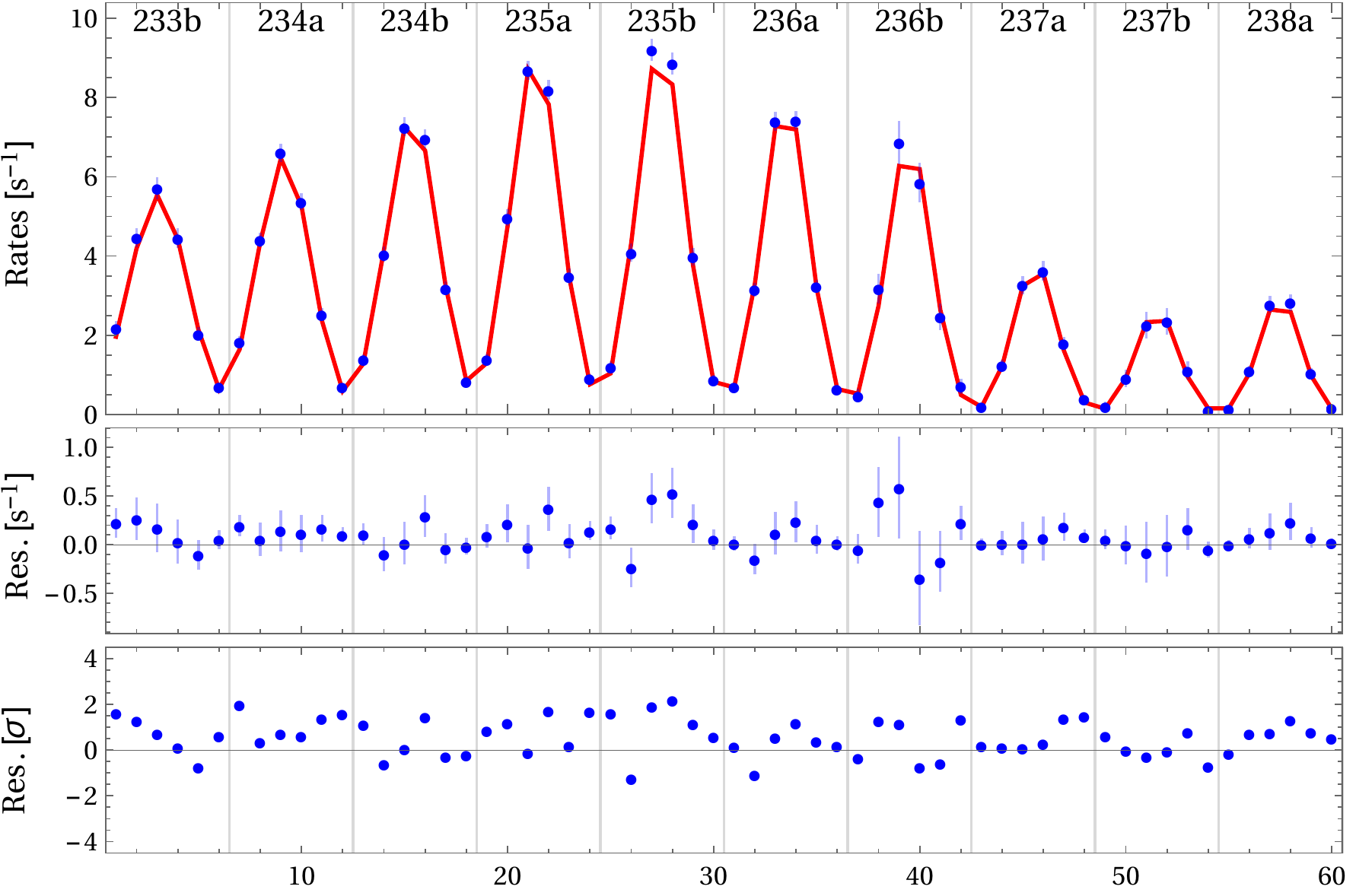}\\
	\end{tabular}
	\caption{Data vs. best-fit simulation for all six ISN~He seasons together, obtained from the direct-fitting method. In each of the panels, the top plot presents the simulated signal (red line) and the data with their error bars, which are small in the scale of the figure (blue points). The orbit numbers are listed at the upper horizontal axis. In the lower horizontal axis, the data point number during the season is given. The middle subpanel presents the absolute difference between the data and the simulation (with absolute uncertainties), scaled in counts/second/bin. In the lower subpanel, these differences are scaled by their respective uncertainties.}
	\label{fig:allYearsDirResids}
	\end{figure}

The ISN parameter uncertainties reported in the first paragraph of this section are taken directly from the covariance matrix. Based on the covariance matrix, one could also split the uncertainties into uncertainty parts along and across the correlation tube, as described by \citet{swaczyna_etal:15a} (in particular, see their Equations~22 and 23). Finally, the correlation line, superimposed on the cuts through the calculation grid in the parameter space, is shown in Figure~\ref{fig:allYearsDirCorLine}. Note that the temperatures in the figures and in the matrices from Equations~\ref{eq:allYearsDirCorr} and \ref{eq:allYearsDirCov} are given in thousands of Kelvins. The comparison of the data and simulations, and the residuals in two different views, are shown in Figure~\ref{fig:allYearsDirResids}. 

As pointed out by \citet{swaczyna_etal:15a}, there is a very high correlation between temperature and speed: the correlation coefficient is equal to $\sim0.95$.  Also, a very strong correlation exists between the temperature and longitude, and speed and longitude. This manifests itself in the existence of a ``correlation tube'' in the four-dimensional parameter space, as described above and shown in three cuts in Figure~\ref{fig:allYearsDirCorLine}. The minimum value of $\chi^2 = \sim 467$ found is very high compared with the expected value, which is equal to the number of degrees of freedom $N_{\mathrm{dof}} = 254$. It is larger by more than 9 standard deviations (equal to $\sqrt{2 N_{\mathrm{dof}}}$) than the expected value. This suggests that a component in the signal still remains unaccounted for in the analysis, or that the physical model adopted here is not fully adequate, or that the uncertainty system is not complete. This latter possibility makes us scale up the uncertainties obtained formally from the covariance matrix. 

A formally unacceptably high value of the minimum $\chi^2$ obtained from the fitting procedure is not an unusual situation in physics and astronomy. One of the reasons for the high value of $\chi^2$ could be that despite all of our efforts, the uncertainties are still underestimated. In that case, the simplest procedure to bring the $\chi^2$ value into an acceptable range is to scale the measurement uncertainties by a common factor equal to the square root of the obtained minimum $\chi^2$ value divided by the number of degrees of freedom \citep{olive_etal:14a}: 		
	\begin{equation}
	S=\left( \chi^2 / N_{\mathrm{dof}} \right)^{1/2}.
	\label{eq:defRedChi}
	\end{equation}
This is equivalent to multiplying the data covariance matrix by $S^2$. This procedure does not change the position of the $\chi^2$ minimum (i.e., the best-fit parameter set), but now the minimum $\chi^2$ value is exactly equal to the number of degrees of freedom, i.e., to the expected value. This leads to the magnification of the obtained covariance matrix of the fitted parameters by the same factor $S^2$, and thus the final uncertainties are also increased by the factor $S$. All of the following uncertainties, including the uncertainties in Tables~\ref{tab:resultCollection} and \ref{tab:resultsYearly}, are obtained following this procedure. A discussion of this error scaling is provided in the discussion section.

The very high correlation between speed and temperature is obtained directly from the fitting process without any additional assumptions. Thus, the existence of the ``tube'' in the 4D parameter space, shown in the original ISN~He analysis by \citet{lee_etal:12a, mobius_etal:12a}, and mostly by \citet{mccomas_etal:12b} based on a simplified analytical model, is experimentally confirmed using a full model of the ISN~He flow. Its presence suggests that, effectively, the most robust parameter that can be derived from this analysis is the Mach number of the ISN~He flow in front of the heliosphere:	
	\begin{equation}
	M_{\mathrm{ISNHe}}=\frac{v_{\mathrm{B}}}{\sqrt{\case{5}{3} \frac{k_{\mathrm{B}} T}{m_{\mathrm{He}}}}},
	\label{eq:defMachNumber}
	\end{equation}
where $v_{\mathrm{B}}$ is the bulk speed of ISN~He in front of the heliosphere, $m_{\mathrm{He}}$ is the mass of He atoms, $T$ the ISN~He temperature, and $k_{\mathrm{B}}$ is the Boltzmann constant. The ISN~He Mach number obtained from the fit to the entire data set is equal to $5.079 \pm 0.028$, where the uncertainty was calculated using the full covariance matrix obtained in the fitting, listed in Equation~\ref{eq:allYearsDirCov}, and scaled by the factor $S$ obtained from Equation~\ref{eq:defRedChi}. 
	
To further verify the robustness of the result, we carried out a complementary analysis using the Gaussian parameter fitting method described above, which is closely related to the method used by \citet{mobius_etal:12a}, \citet{leonard_etal:15a}, and \citet{mobius_etal:15b}. The results obtained from this method are in close agreement with the global fit discussed above ($\lambda = 256.1\degr \pm 0.6$, $\beta = 5.13\degr \pm 0.12$, $T = 7330 \pm 350$~K, $v = 25.6 \pm 0.5$~\kms, and $M_{\mathrm{ISNHe}} = 5.074 \pm 0.034$), and the conclusions about the presence and strength of the correlation between the parameter pairs are supported. Again, the minimum value of $\chi^2$ is well above the expected value, even more so than for the direct-fitting method. The results from the direct-fitting and Gaussian parameter fitting methods are within their respective $1\sigma$ uncertainties. They are collected in Table~\ref{tab:resultCollection}. In addition to the results just discussed, this table includes results from fits to a subset of data that will be explored in Section \ref{sec:dependondatasel}. 

\begin{deluxetable}{lcccccccc}
\tabletypesize{\scriptsize}
\tablecaption{Parameter fitting results for the full and L$\&$M data sets, obtained using direct-fit and Gaussian parameter fit methods, and the calculated thermal Mach numbers for the ISN~He flow.\label{tab:resultCollection}}
\tablewidth{0pt}
\tablehead{
\colhead{Method / data set} & \colhead{$\lambda \left[\degr\right]$} & \colhead{$\beta \left[\degr\right]$} & \colhead{$T \left[\mathrm{K}\right]$} & \colhead{$v \left[\mathrm{km~s}^{-1}\right]$} & \colhead{$\chi^2$} & \colhead{$N_{\mathrm{dof}}$} & \colhead{$\Delta\left[\sigma\right]^{a}$} & \colhead{$M_{\mathrm{ISNHe}}$}
}
\startdata
Direct fit, full set & 255.8 $\pm$  0.5 & 5.16 $\pm$ 0.10 & 7440 $\pm$ 260 & 25.8 $\pm$ 0.4 & 467.3 & 254 & 9.5 & 5.079 $\pm$ 0.028 \\
Gauss fit, full set & 256.1 $\pm$ 0.6 & 5.13 $\pm$ 0.12 & 7330 $\pm$ 350 & 25.6 $\pm$ 0.5 & 395.6 & 122 & 17.5 & 5.074 $\pm$ 0.034 \\
Direct fit, L$\&$M set & 255.3 $\pm$ 0.6 & 5.14 $\pm$ 0.16 & 8150 $\pm$ 390 & 26.7 $\pm$ 0.5 & 125.8 & 83 & 3.3 & 5.029 $\pm$ 0.049 \\
Gauss fit, L$\&$M set & 255.0 $\pm$ 0.7 & 5.09 $\pm$ 0.11 & 8010 $\pm$ 410 & 26.7 $\pm$ 0.6 & 65.8 & 38 & 3.2 & 5.069 $\pm$ 0.039 \\
Ulysses (Bzowski 2014)$^{b}$ & 255.3 $\pm$ 1.2 & 6.0 $\pm$ 1.0 & 7500 $\pm$ 1500 & 26.0 $\pm$ 1.5 & & & &5.10\\
Ulysses (Wood 2015)$^{c}$ & 255.54 $\pm$ 0.19 & 5.44 $\pm$ 0.24 & 7260 $\pm$ 270 & 26.08 $\pm$ 0.21 & & & & 5.20 \\
\enddata
\tablenotetext{a}{By $\Delta$ we denote the difference between the obtained $\chi^2$ value and its expected value, equal to the number of degrees of freedom $N_{\mathrm{dof}}$ in the fit, divided by the square root of two times the number of degrees of freedom.}
\tablenotetext{b}{From \citet{bzowski_etal:14a}.}
\tablenotetext{c}{From \citet{wood_etal:15a}.}
\end{deluxetable}

\begin{figure}
	\centering
	\begin{tabular}{cc}
	\includegraphics[scale=0.4]{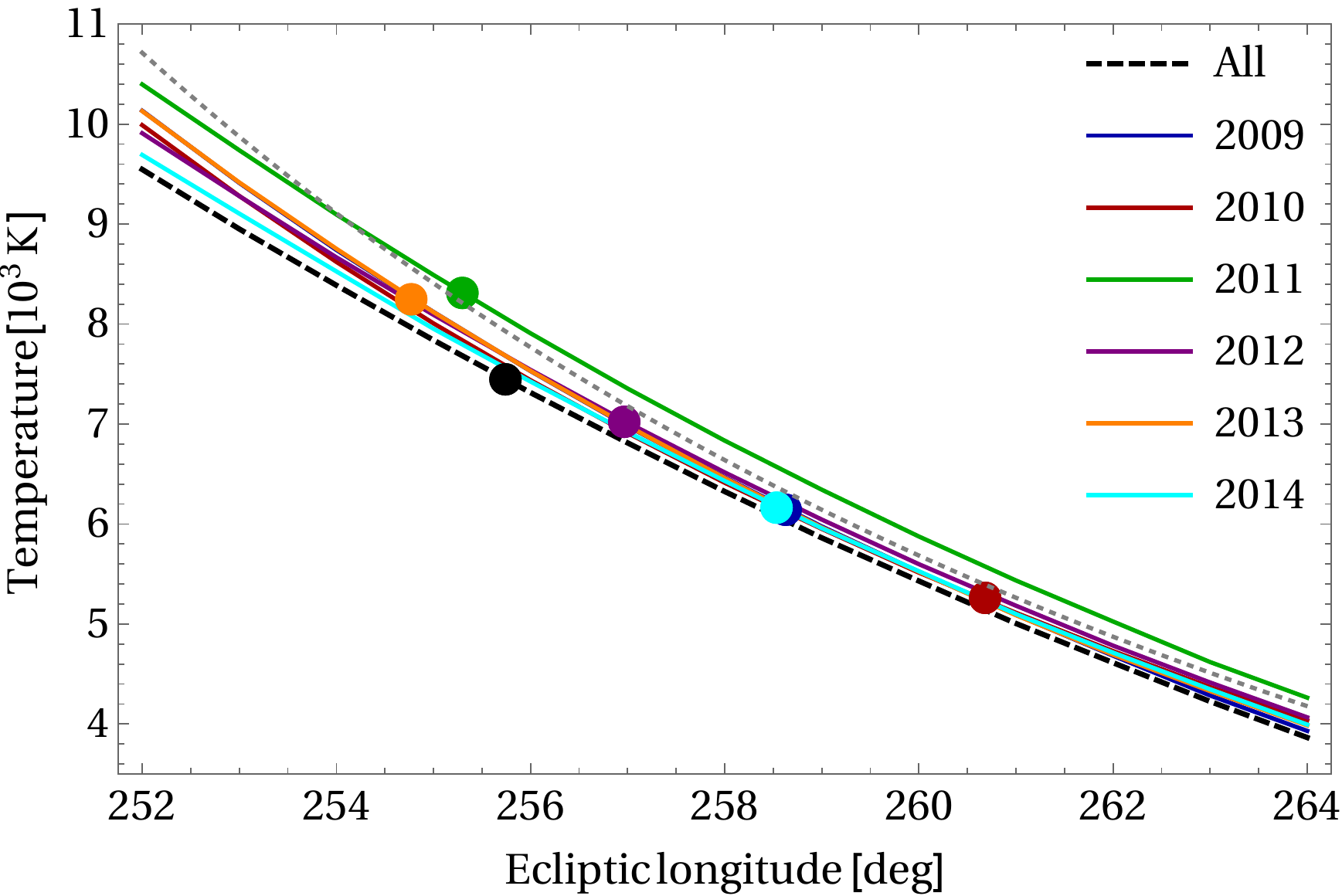} & \includegraphics[scale=0.4]{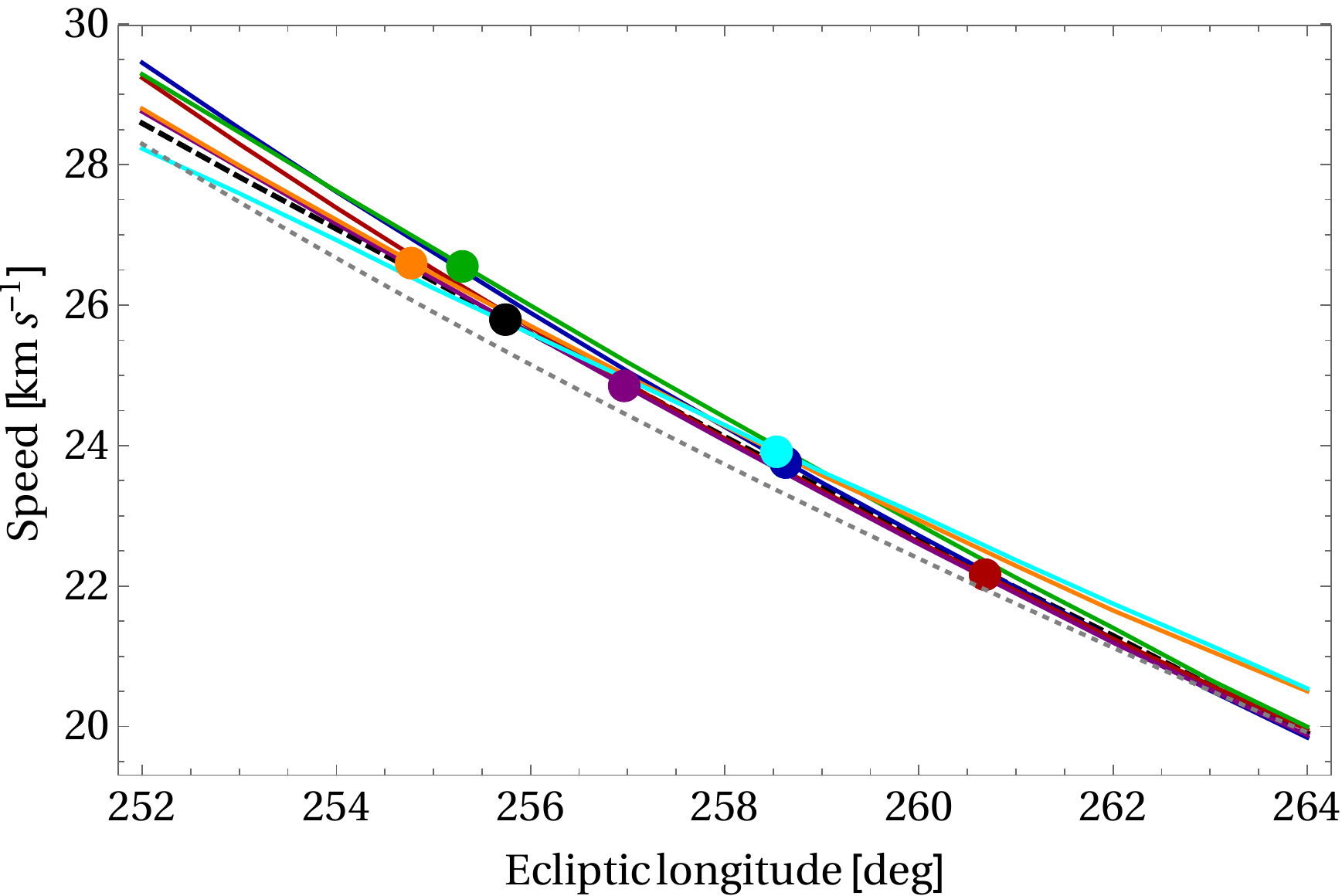}\\
	\includegraphics[scale=0.4]{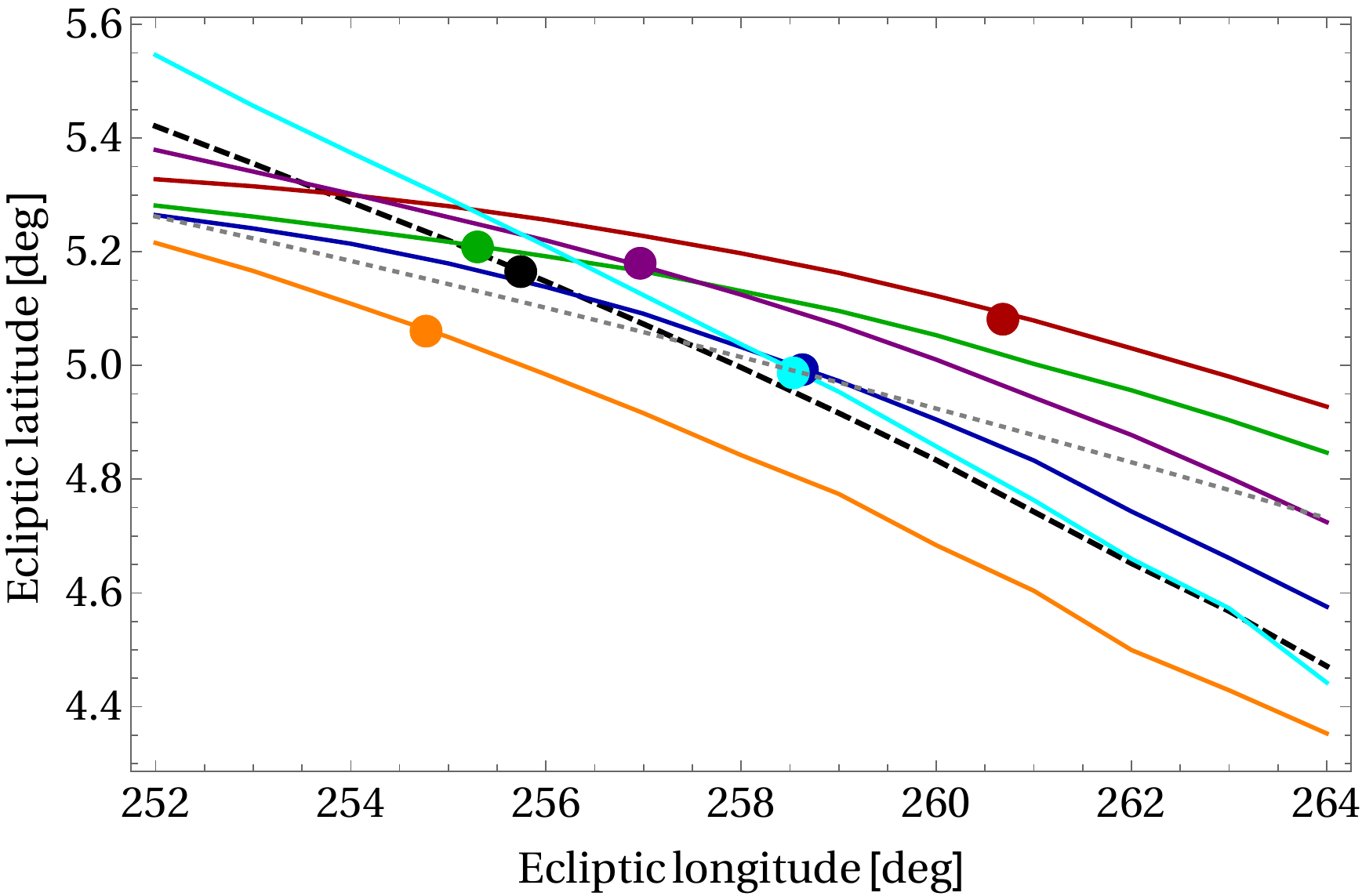} & \includegraphics[scale=0.4]{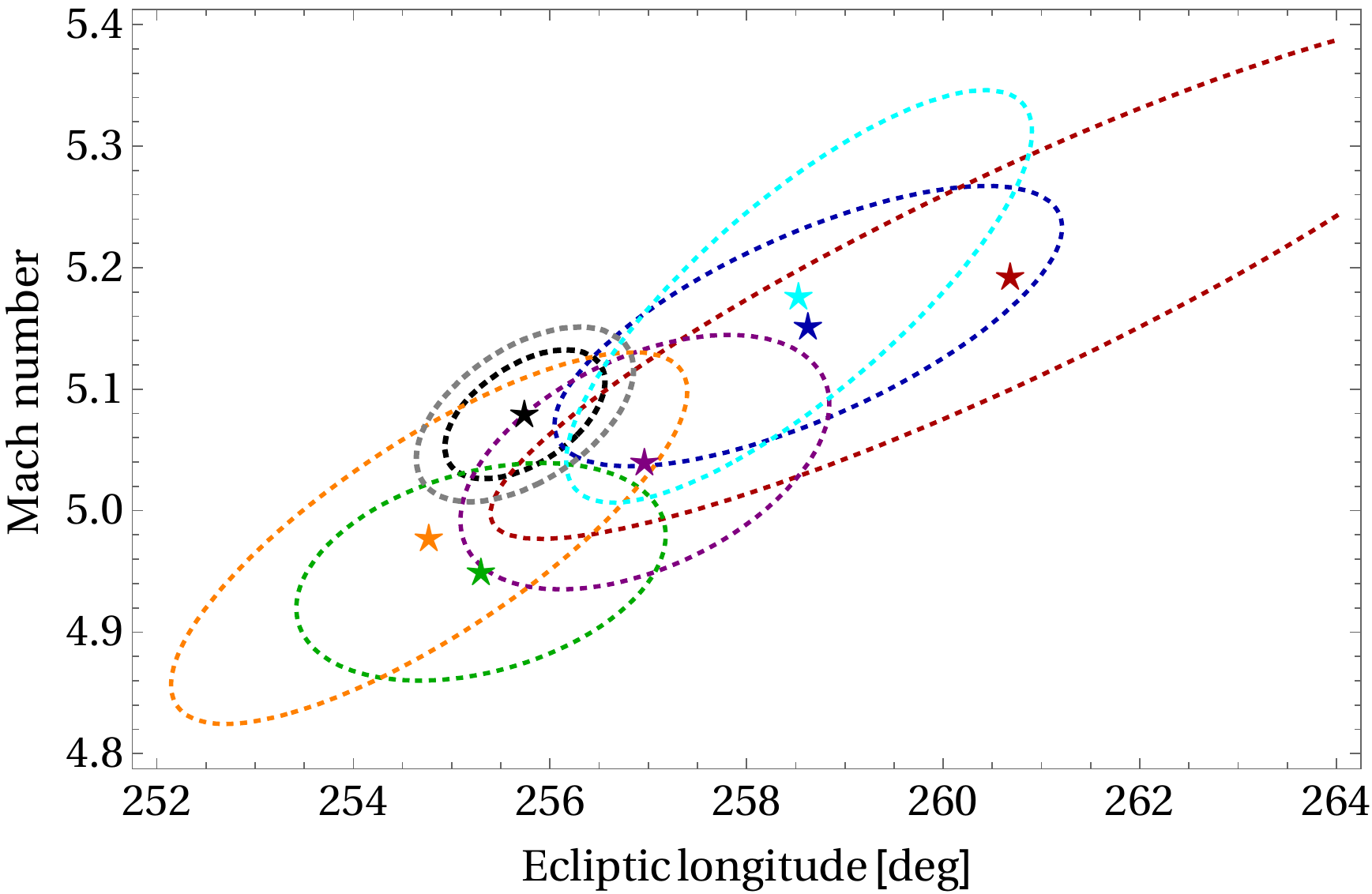}\\
	\end{tabular}
	\caption{Cuts through the lines of $\chi^2\left(\lambda\right)$ in the four-dimensional parameter space. The upper left panel shows the temperature vs. longitude, the upper right panel the speed vs. longitude, and the lower left panel latitude vs. longitude. The dashed line corresponds to the global fit to all of the seasons together. The dotted line presents the parameter tube obtained by \citet{mccomas_etal:12b} based on the results from \citet{mobius_etal:12a} and \citet{bzowski_etal:12a}. The color lines represent the parameter tubes obtained from individual seasons (see the legend in the upper left panel), and the best-fit solutions for these seasons are marked with dots. The lower right-hand panel shows the Mach numbers as a function of inflow longitude, obtained from the fits to individual seasons and from the global fit (stars), and ``$2 \sigma$'' contours for the individual seasons and for the global fit, obtained directly from the fitting. The gray oval is the black contour obtained for the global fit, scaled up as described in the text using Equation \ref{eq:defRedChi} to account for the high minimum $\chi^2$ value.}
	\label{fig:parameterLines}
	\end{figure}

In addition to fitting the data from all of the observation seasons together, we performed fits to data from the individual seasons. The results are collected in Table~\ref{tab:resultsYearly}. The global fit and the fits from individual seasons yield very similar, but not identical correlation lines, as shown in Figure~\ref{fig:parameterLines}, which illustrates cuts through the $\chi^2\left(\lambda\right)$ lines in the four-dimensional parameter space. The correlation lines from individual seasons are very close to each other and to the correlation lines obtained for the entire data set, as well as to the correlation line found by \citet{mccomas_etal:12b}. Additionally, Figure~\ref{fig:parameterLines} presents the scaled $2 \sigma$ covariance ellipsoid for the inflow longitude versus the Mach number of the ISN~He flow, calculated from the inflow speed and temperature obtained from the global fit and from the fits to individual seasons. Clearly, the Mach number obtained from the fits and the fit longitude of the inflow are correlated, as suggested by the correlation matrix and as it emerges from the simplified analytical theory by \citet{lee_etal:12a} (see also \citet{mccomas_etal:12b}. 

\begin{deluxetable}{lcccccccc}
\tabletypesize{\scriptsize}
\tablecaption{ISN~He inflow parameters obtained for individual observation seasons from the direct fitting method and the calculated thermal Mach numbers for ISN~He. 
\label{tab:resultsYearly}}
\tablewidth{0pt}
\tablehead{
\colhead{Year} & \colhead{$\lambda \left[\degr\right]$} & \colhead{$\beta \left[\degr\right]$} & \colhead{$T \left[\mathrm{K}\right]$} & \colhead{$v \left[\mathrm{km~s}^{-1}\right]$} &
\colhead{$\chi^2$} & \colhead{$N_{\mathrm{dof}}$} & \colhead{$\Delta\left[\sigma\right]^{a}$} & \colhead{$M_{\mathrm{ISNHe}}$}
}
\startdata
2009 & 258.6 $\pm$ 1.2 & 4.99 $\pm$ 0.16 & 6140 $\pm$ 590 & 23.8 $\pm$ 1.0 & 44.3 & 31 & 1.7 & 5.152 $\pm$ 0.055 \\
2010 & 260.7 $\pm$ 2.7 & 5.08 $\pm$ 0.22 & 5260 $\pm$ 1130 & 22.2 $\pm$ 2.0 & 40.6 & 25 & 2.2 & 5.193 $\pm$ 0.104 \\
2011 & 255.3 $\pm$ 0.9 & 5.21 $\pm$ 0.18 & 8310 $\pm$ 580 & 26.6 $\pm$ 0.8 & 38.0 & 25 & 1.8 & 4.950 $\pm$ 0.044 \\
2012 & 257.0 $\pm$ 1.3 & 5.18 $\pm$ 0.23 & 7020 $\pm$ 680 & 24.9 $\pm$ 1.0 & 123.6 & 43 & 8.7 & 5.040 $\pm$ 0.071 \\
2013 & 254.8 $\pm$ 1.4 & 5.06 $\pm$ 0.22 & 8250 $\pm$ 900 & 26.6 $\pm$ 1.1 & 95.2 & 55 & 3.8 & 4.977 $\pm$ 0.082 \\
2014 & 258.5 $\pm$ 1.0 & 4.99 $\pm$ 0.17 & 6160 $\pm$ 470 & 23.9 $\pm$ 0.7 & 54.0 & 55 & -0.1 & 5.176 $\pm$ 0.089 \\
\enddata
\tablenotetext{a}{By $\Delta$ we denote the difference between the obtained $\chi^2$ value and its expected value, equal to the number of degrees of freedom $N_{\mathrm{dof}}$ in the fit, divided by the square root of two times the number of degrees of freedom.}
\end{deluxetable} 

The inflow longitudes from individual seasons are in agreement with each other at the $2\sigma$ level of their individual uncertainties, listed in Table~\ref{tab:resultsYearly}, but outside the $1\sigma$ uncertainties of the result obtained for the combined data set. Regardless of the correlation with the fit inflow longitude, the Mach numbers obtained for individual seasons and from the global analysis are in agreement within $\sim2.5\%$, which is much better than the spread obtained in the temperature and the components of the inflow speed vector. Thus, it can be regarded as the most tightly constrained parameter of the ISN~He inflow into the heliosphere observed by IBEX. 
	
\section{Discussion}
\label{sec:Discussion}
In the discussion we consider the following related issues: (1) does the result depend on the data selection, and (2) how do the results compare for analyses of the same data using different methods? We also discuss the question of what the high minimum $\chi^2$ value suggests and point out some implications of the obtained results.

\subsection{Do the results from individual seasons statistically agree with the result from the global fit?}
\label{sec:dotheyagree}
The inflow longitudes obtained from the fits to all of the individual seasons do not form any statistical trend with time. To check whether results from individual seasons are statistically consistent with the result from the global fit, we check if differences between the parameters from a single season $\vec{\pi}_k = \left(\lambda_k, \beta_k, T_k, v_k\right)$ and from the global fit $\vec{\pi}_{\mathrm{ISNHe}} = \left(\lambda_{\mathrm{ISNHe}}, \beta_{\mathrm{ISNHe}}, T_{\mathrm{ISNHe}}, v_{\mathrm{ISNHe}}\right)$ are statistically consistent with zero. The fit results are vectors in the parameter space and we need to define the distance between two points in this space, taking into account the uncertainties of the vector positions and correlations between the vector elements. For this distance, we adopt the following definition, which takes into account the uncertainties and correlations between the parameters obtained from the fits:		
	\begin{equation}
	\chi_{k}^{2} = \left( \vec{\pi}_k - \vec{\pi}_{\mathrm{ISNHe}} \right)^{\mathrm{T}} \cdot \left( \mathbf{\Sigma}_{k} S_{k}^{2} + \mathbf{\Sigma}_{\mathrm{ISNHe}} S_{\mathrm{ISNHe}}^2 \right)^{-1} \cdot \left( \vec{\pi}_k - \vec{\pi}_{\mathrm{ISNHe}} \right)
	\label{eq:chi2consistency}
	\end{equation}
where $k$ enumerates the individual seasons, $\mathbf{\Sigma}_{k}$ is the covariance matrix obtained from the fit to the $k$-th season, $\mathbf{\Sigma}_{\mathrm{ISNHe}}$ is the covariance matrix for the global fit (Equation~\ref{eq:allYearsDirCorr}), and $S_k$, $S_{\mathrm{ISNHe}}$ are the scaling factors calculated from Equation~\ref{eq:defRedChi} for the $k$-th season and the global fit, respectively. The covariance matrix for the difference between the parameters obtained for an individual season and the global fits is calculated by uncertainty propagation given by the sum of the respective covariance matrices. Note that we scale our obtained covariance matrix by the respective factors $S_k$, $S_{\mathrm{ISNHe}}$, to acknowledge the increased uncertainty due to the high value of the minimum $\chi^2$ for the fits, and that we assume that the covariance matrix for an individual season $\mathbf{\Sigma}_k$ is independent of the covariance matrix for the global fit $\mathbf{\Sigma}_{\mathrm{ISNHe}}$. Since the matrix $\mathbf{\Sigma}_k S_{\mathrm{ISN}}^2$  in all entries is much smaller than $\mathbf{\Sigma}_k S_{k}^2$, this assumption should not significantly affect the result. The matrix $\mathbf{\Sigma}_{\mathrm{ISNHe}}$ is given by Equation~\ref{eq:allYearsDirCov}, whereas matrices $\mathbf{\Sigma}_{k}$ are part of the {\it IBEX} data release \citep{schwadron_etal:15a, swaczyna_etal:15a, sokol_etal:15b}.
	\begin{figure}
	\centering
	\resizebox{\hsize}{!}{\includegraphics{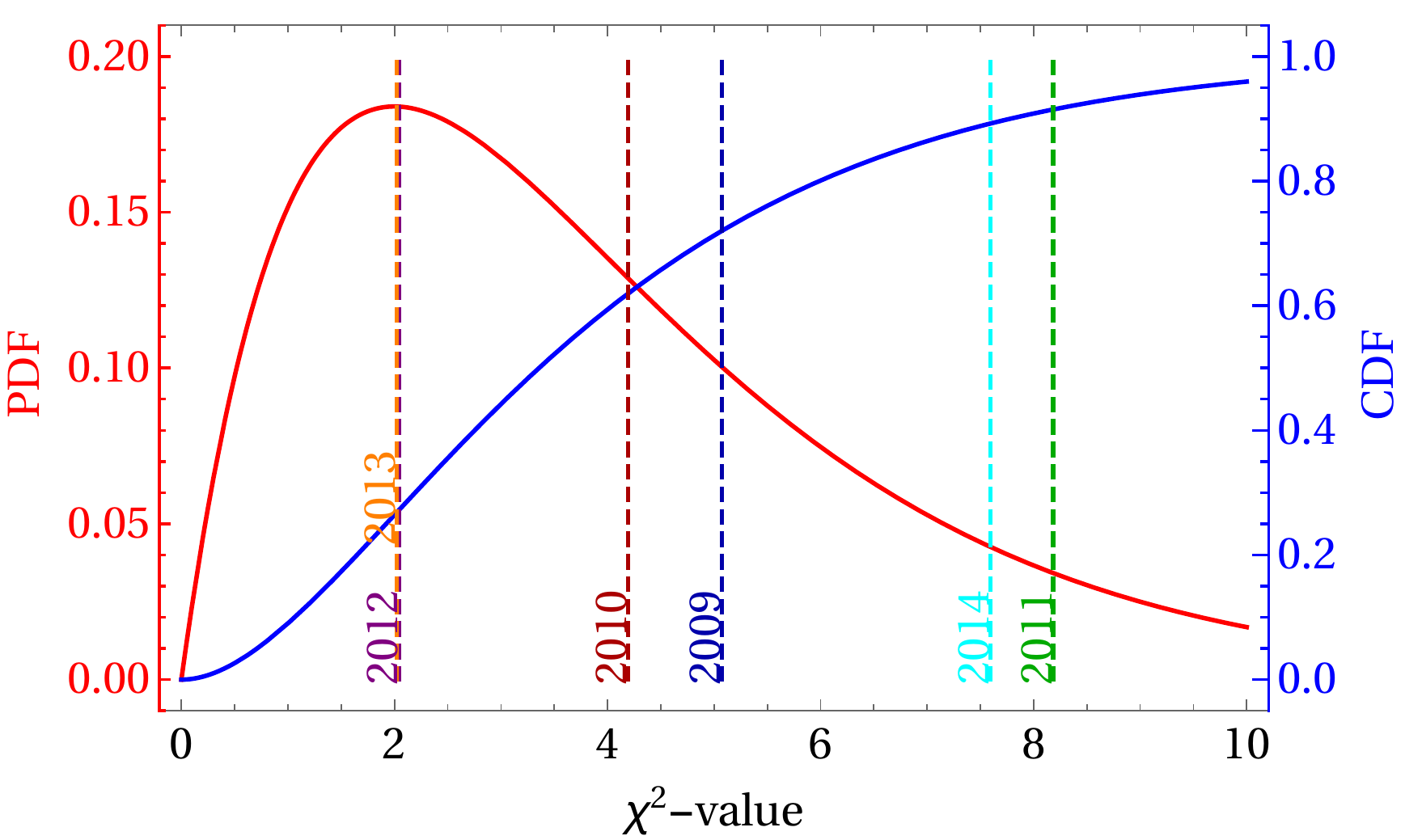}}
	\caption{$\chi^2$ values given by Equation~\ref{eq:chi2consistency} for fits to the data from individual seasons, denoted by the dashed vertical lines, overlaid on the plots of the probability distribution function (PDF, left-hand vertical scale) and cumulative distribution function (CDF, right-hand vertical scale) for the $\chi^2$ probability distribution function for 4 degrees of freedom. Note that the results for 2012 and 2013 are very close to each other (almost merged in the plot).}
	\label{fig:chi2consistency}
	\end{figure}
	
The quantity returned by Equation~\ref{eq:chi2consistency}, i.e., the ``distance'' between the global solution $\vec{\pi}_{\mathrm{ISNHe}}$ and the solution $\vec{\pi}_{k}$ from an individual season $k$, should be distributed as the $\chi^2$ probability distribution with 4 degrees of freedom because we have 4 parameters, and the statistical model tested (the consistency with zero) has no free parameters. The values obtained for seasons 2009--2014 are presented in Figure~\ref{fig:chi2consistency}, where we overlay them on a plot of the probability distribution function (PDF) and the cumulative distribution function (CDF) of the $\chi^2$ probability distribution for 4 degrees of freedom. Note that the $\chi^2$ distribution for the low number of degrees of freedom does not resemble the Gaussian (normal) distribution: it is asymmetric relative to the peak (most probable) value, and the expected value (mean), equal to the number of degrees of freedom, is not equal to the most probable value. The largest value is obtained for season 2011, for which we obtain a ``distance'' equal to 8.22. The smallest ``distance'' from the global fit was obtained for 2013, i.e., for the season with the best temporal coverage. This is because 2013 is the season that has the largest influence on the final result among the individual seasons, even though the minimum $\chi^2$ obtained from fitting this season deviates by $3.8 \sigma$ from the expected value (see Table~\ref{tab:resultsYearly}).

We treat the ``distances'' between the results for individual seasons and the global fit, obtained from Equation~\ref{eq:chi2consistency}, as random draws from a population that is expected to follow the chi-squared distribution with 4 degrees of freedom. We have a series of six such draws (and thus six values of ``distances'' $\chi_k^2$, $k = 2009,...,2014$, corresponding to the $\chi^2$ axis in Figure~\ref{fig:chi2consistency}). We ask the following question. If $\chi_k^2$ is randomly drawn from a population following the $\chi^2$ distribution with 4 degrees of freedom, then what is the probability $\alpha_k$ that such a random draw returns a value not larger than this given $\chi^2_k$? We follow the common scheme for finding the significance level of agreement of an experimental result with expectations. The probability $\alpha_k$ is given by the cumulative distribution function (CDF) of the $\chi^2$ distribution with 4 degrees of freedom. For $\alpha = 0.68$ (i.e., ``$1 \sigma$''), $\chi^2_k$ would need to be $\leq 4.75$. For $\alpha = 0.955$ (`$2 \sigma$''), $\chi^2_k$ would need to be  $\leq 9.7$. Thus, our largest sampled $\chi^2_{2011} = 8.22$ agrees with the global result to better than $2 \sigma$, but worse than $1 \sigma$.

Additionally, we calculated the sum of the $\chi^2_k$ values from individual seasons, which is equal to 29.03, and checked the likelihood of the hypothesis that it is randomly drawn from a $\chi^2$ probability distribution with 24 degrees of freedom (6 seasons with 4 parameters each). In that case, the CDF value for an argument of 29.03 is equal to 0.78. Thus, this result also confirms the consistency of the results from individual seasons. Another simple test involves calculating the arithmetic mean value of the $\chi^2_k$ values from the six individual seasons and comparing it with the expected value of 4. This arithmetic mean is equal to $4.8 \pm 1.2$, in a very good agreement with expectations. Thus, we conclude that the overall scatter of the ISN~He parameters obtained from fits to data from individual seasons around the parameter set obtained from the fit to the full data set is in good agreement with the statistical expectations.

\subsection{Does the global fit result depend on data selection?}
\label{sec:dependondatasel}
The results obtained here for the full data set are very similar to the results we obtained using the data selection proposed process by \citet{leonard_etal:15a} and \citet{mobius_etal:15b}, which we denote as the L$\&$M selection. \citet{leonard_etal:15a} and \citet{mobius_etal:15b} selected those orbits for which the inclination of the spin axis to the ecliptic plane was less than $0.2\degr$. This included a subset of orbits only from seasons 2012 to 2014, with the orbits from season 2013 being the most numerous. Adopting the L$\&$M orbit selection and the spin-angle range restricted to $252\degr-282\degr$, identical to the range used in our global analysis, results in best-fitting ISN~He parameters of $\lambda = 255.3\degr$, $\beta = 5.14\degr$, $T = 8150$~K, $v = 26.7$~\kms, and $M_{\mathrm{ISNHe}} = 5.028$ using the direct-fitting method (for uncertainties, see Table~\ref{tab:resultCollection}).

The parameters obtained from this restricted set are also expected to agree within the statistical uncertainty with the parameters obtained from the full set, and an approximate agreement has indeed been obtained (see the former subsection for the agreement criteria). The ``distance'' from the global best-fit $\chi^2_{\mathrm{L}\&\mathrm{M}} = 4.91$, calculated from Equation~\ref{eq:chi2consistency}, is very close to the expected value of 4. Broadening the spin-angle selection by two pixels each way, i.e., adopting the spin-angle range $240\degr-294\degr$, which implies an extension into the range where the potential influence of the Warm Breeze increases, does not significantly change the result statistically: $\lambda = 255.8\degr$, $\beta = 5.13\degr$, $T = 7790$~K, $v = 26.3$~\kms, $M_{\mathrm{ISNHe}} = 5.057$. Also, using the Gaussian parameter fit method instead of direct flux fitting provides similar results ($\lambda = 255.0\degr$, $\beta = 5.08\degr$, $T = 8030$~K, $v = 26.7$~\kms, $M_{\mathrm{ISNHe}} = 5.062$), and the $\chi^2$ distance of this result from the global fit is 2.92. The uncertainties of these parameters are listed in Table~\ref{tab:resultCollection}. The L$\&$M selection includes approximately one-third of the points we used in the global fit. The $\chi^2$ values obtained for the L$\&$M selection are also too high, but their departures from the expected value is only $\sim3\sigma$. This is likely because the statistics are lower by a factor of three. Independent of \citet{leonard_etal:15a}, the L$\&$M subset was also analyzed by \citet{schwadron_etal:15a}, who obtained very similar results: $\lambda = 255.8\degr$, $\beta = 5.11\degr$, $T = 7900$~K, and $v = 25.4$~\kms. For this subset, \citet{leonard_etal:15a} report $\lambda = 254.5\degr \pm 1.7\degr$, $\beta = 5.2\degr \pm 0.3\degr$, $v = 27.0 +1.4/-1.3$~\kms, and \citet{mobius_etal:15b} supplement this with a temperature $T = 8650 +450/-740$~K.  Thus, these results seem quite robust both from the viewpoint of the expected statistical scatter within our uncertainty system, and from the viewpoint of the application of different, independent analysis methods. 

The internal consistency of the result and modeling is further supported by our analysis of the scaling factors obtained from the fitting, listed in the Results section. These factors should be considered in pairs. The factors for the first two seasons are treated as the baseline. The ionization rate during these years varied little and both the data collection process and the instrument settings were unmodified. The scaling factors obtained for 2009 and 2010 are almost identical. During 2011 and 2012, the instrument was operated in a special mode and the data used for the instrument interface throttling correction, presented by \citet{swaczyna_etal:15a}, could not be obtained with a sufficient resolution. Thus, we do not have the exact corrections for these years, but assuming that the ambient electron signal, responsible for the throttling, was close to the signal observed during 2009 and 2010, we must rescale the two factors upward by 10--15\%. With this correction, the 2011 and 2012 scaling factors are in excellent agreement with those for 2009 and 2010. In 2013 and 2014, there is no throttling effect but the post-acceleration voltage in the instrument had to be lowered, which resulted in an expected reduction of the instrument efficiency by a factor of 0.44. When applied to the correction factors for 2013 and 2014, this brings them to the values within $\pm 5$\% of those for 2009 and 2010. This check suggests the conclusions that (1) there was no change in the instrument sensitivity unaccounted for in our analysis system during the six analyzed seasons, and (2) our ionization rate model and the system of calculation of the ionization losses is also correct. Thus, we do not need to expect any bias due to the unknown instrument sensitivity changes and inaccurate variations in the ionization losses in our results. 

\subsection{Possible reasons and implications of the overly high minimum $\chi^2$ value from the global fit}
\begin{figure}
	\centering
	\resizebox{\hsize}{!}{\includegraphics{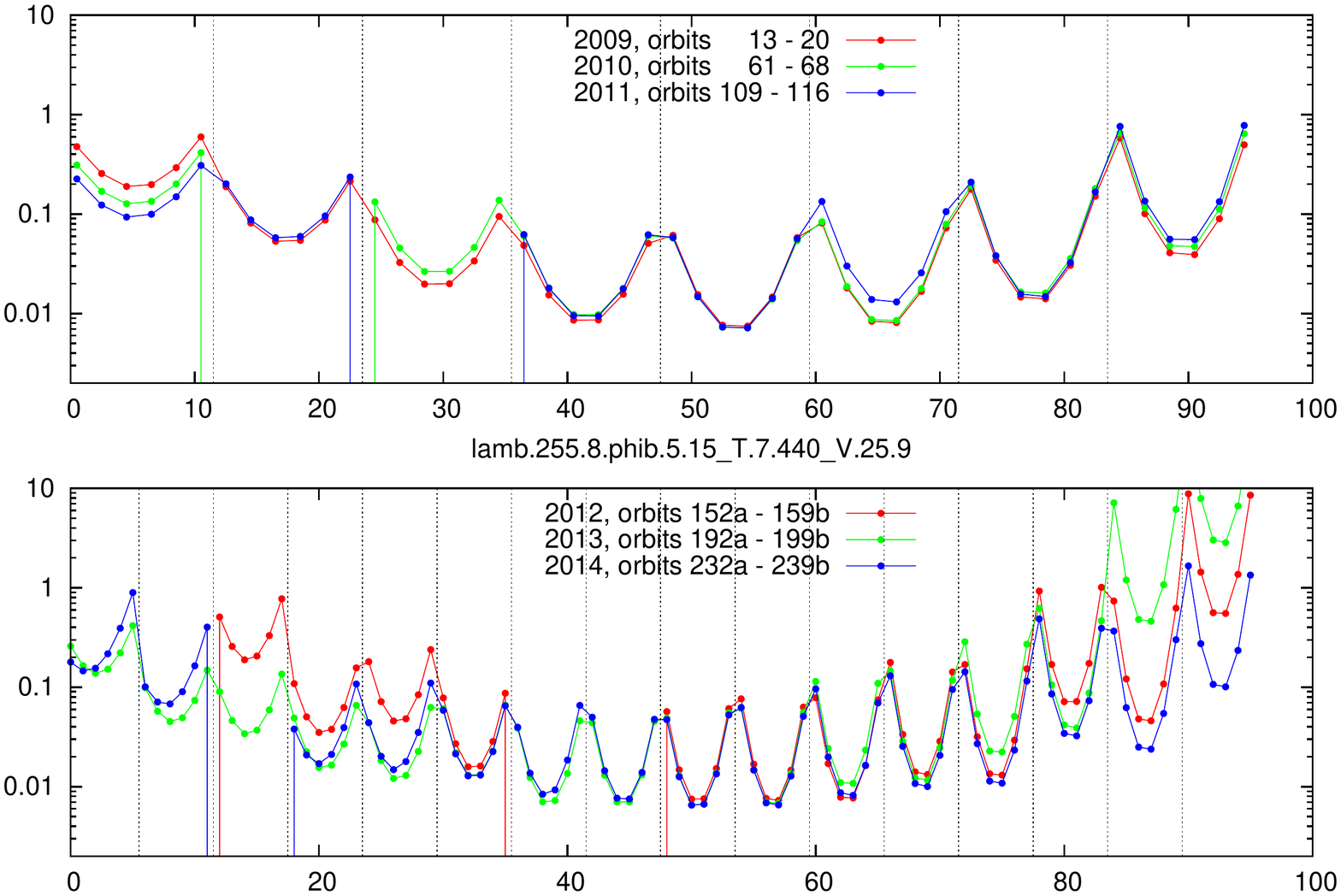}}\\
	\caption{Ratio of the Warm Breeze flux to the flux from ISN~He simulated for the parameters obtained using the direct-fitting method for all six ISN~He seasons. The comparison of the six ISN~He seasons is split into two panels, with the seasons marked by the color code explained in the panels. Vertical bars separate the orbits. The spin-angle range shown corresponds to the $\left(252\degr,\quad 282\degr\right)$ range used in the parameter fitting. The orbits shown include the range used in the fitting, plus one orbit more toward lower and higher ecliptic longitudes. The horizontal axis represents the sequence of the data points for each individual season. Note that the equivalence between orbits from different seasons was maintained, which results in the visible breaks in some lines that mark the missing data. The  numbering of points in the sequence is adjusted so that the corresponding points during each season are attributed similar numbers. The yearly sets of orbits shown cover nearly identical ranges of ecliptic longitudes.}
	\label{fig:WarmBreezeContri}
	\end{figure}

The minimum $\chi^2$ value obtained from the global fit significantly exceeds the expected value. This is likely due to the fact that the adopted physical model is missing some important aspect and/or element. One of the highly uncertain elements in the model is the Warm Breeze. The contribution from the Warm Breeze to the signal is not negligible and its influence on the fitting result is apparently substantial. The ratio of the Warm Breeze flux to the ISN~He flux, calculated for the observation conditions for all six ISN~He seasons and an ISN~He parameter set close to the best-fit solution, is presented in Figure~\ref{fig:WarmBreezeContri}. Since the temperature of the Warm Breeze found by \citet{kubiak_etal:14a} is at least two times greater than the temperature of ISN~He, it is not surprising that its contribution at the early and late orbits during the season is the largest, and for a given orbit, the spin-angles farthest from the ISN~He peak are the most heavily affected. It is interesting, however, that even though the inflow longitude of the Warm Breeze is lower than the inflow longitude of ISN~He, the early orbits in each season do not appear to be more affected than the late ones.

The relative contributions from the Warm Breeze to the simulated signal are the lowest and the least variable at the peak orbits in each season, and are the largest in the data points farthest from the orbital maximum. Beyond that, however, they do not show any systematic trend: there is no strong dependence on the ionization level, which was increasing from 2009 to 2014 \citep{sokol_bzowski:14a}, there is no individual season when the Breeze contribution is consistently the largest, etc. The fact that the relative contributions vary from orbit to orbit and from one spin angle bin to another is likely due to details such as the spin axis pointing, {\it IBEX} orbital motion, and length and placement of ISN good time intervals during the orbits. 

Following \citet{swaczyna_etal:15a}, we verified that not subtracting the Warm Breeze from the data significantly worsens the fit quality obtained from the direct-fitting method and slightly shifts the correlation tube in parameter space, but the fitted ISN~He parameters change relatively little (the inflow longitude is still less than $256\degr$). However, using the Gaussian parameter fitting method with the Warm Breeze not subtracted from the data results in an inflow longitude of almost $257\degr$ (with the other parameters changed accordingly), and in a $\chi^2$ value of 609, while the expected value is 118. 

Given the strength of the Warm Breeze signal, it is clear that the accuracy of its model in ISN~He parameter fitting is critical. However, \citet{kubiak_etal:14a} were able to provide only a wide uncertainty range for its parameters: the intensity was uncertain by $\sim50\%$, temperature and speed by $\sim40\%$, and the inflow direction by $12\degr$. These uncertainties were taken into account in our uncertainty system, but if the true values depart from the nominal values suggested by \citet{kubiak_etal:14a}, even within the uncertainty range, then we expect a residual contribution left in the data with the Warm Breeze subtracted, which will affect the present analysis. The Warm Breeze had been fit only to the data from the 2010 ISN~He season, with a simplified uncertainty system and without the insights now available (e.g., the exact value of the {\it IBEX-Lo} energy cutoff for neutral He, obtained by \citet{sokol_etal:15a, galli_etal:15a}). Also, two important assumptions for that analysis were that (1) the primary ISN~He parameters were exactly those found by \citet{bzowski_etal:12a} to be the most likely parameters, and (2) that the parent population for the Warm Breeze is a homogeneous Maxwell--Boltzmann distribution. In fact, neither of these may actually be the case. Given the insight illustrated in Figure~\ref{fig:WarmBreezeContri}, the influence of this hypothetical residual on the fitting results may vary from one observation season to another. In addition, insight provided by \citet{fuselier_etal:14a} and \citet{galli_etal:14a} suggests that there may be a small but non-negligible contribution to the Warm Breeze signal from a local magnetospheric foreground signal, which potentially may be eliminated by comparing the measurements from the three lowest {\it IBEX-Lo} energy channels. Therefore, we are not surprised to see evidence in the present ISN~He analysis that the Warm Breeze deserves to be revisited, with all seasons analyzed simultaneously. This analysis is ongoing, but given the unknown nature of the Warm Breeze on one hand and the complex observational aspects on the other, it is even more challenging than the analysis of the primary ISN He population.

Figure~\ref{fig:allYearsDirResids} provides insights on the contributions of the Warm Breeze . Most of the residuals are positive, even though in a perfect case the mean residual should be 0. Such a behavior is allowed in the uncertainty system from \citet{swaczyna_etal:15a} because of the correlations between the data points. A strong correlating factor in this context is the model of the Warm Breeze that is subtracted from the data. It is striking that the largest excess in the residuals tends to occur during the early orbits in all seasons, where the contribution from the Warm Breeze is largest and, simultaneously, the signal from the Warm Breeze strongly varies from season to season, unlike in the late orbits where the Warm Breeze contribution is also strong, but the season to season variation is much weaker (see Figure~\ref{fig:WarmBreezeContri}). Perhaps the overall density of the Warm Breeze relative to the density of ISN~He found by \citet{kubiak_etal:14a} was a little too high and the fitting system ``compensated'' for it by choosing a solution with a slightly reduced density in the region of the data where the contribution from the Warm Breeze is the largest during the first three seasons (orbits 14 and 15 and equivalent from 2010 to 2011) and relatively less variable. During the 2013 and 2014 seasons, the Warm Breeze contribution in these orbits is lower, which apparently tipped the scales in the global fit toward lower longitudes. Even though the results from three individual seasons suggest a longitude in the range of $258\degr-261\degr$ and the other three other suggest a longitude of $\sim255\degr$ (see Table~\ref{tab:resultsYearly}), duebto the high number of data points in these latter orbits (with two arcs per orbit, and thus twice as many data points), the global fit returned a longitude similar to that from the three seasons of high statistical weight. 

Another potential cause for the high minimum $\chi^2$ value may be a departure of the parent ISN He distribution function toward a kappa function. A kappa distribution with a small $\kappa$ value (i.e., a highly unequilibrated distribution) seems to be ruled out by the analysis performed by \citet{sokol_etal:15a}. However, without a detailed analysis, it cannot be excluded that the parent ISN He distribution is better described by a kappa function with a relatively large $\kappa$ value. \citet{sokol_etal:15a} identify the region in the sky observed by {\it IBEX} where one should look for evidence of such a departure. We plan to verify this hypothesis once the Warm Breeze contribution to the signal is better understood. however, even if the parent distribution is indeed somewhat kappa-like, i.e., it features elevated tails, it is unlikely that the inferred bulk velocity vector would be biased because the symmetry is not changed. It is likely, however, that such a hypothetical departure could increase the minimum $\chi^2$ value if a fit to the Maxwell--Boltzman function is attempted. Similarly, additional filtration in the outer heliosheath due to an increased charge exchange rate would modify the signal observed by IBEX. If relatively strong and not axially symmetrical, then it could bias the obtained flow vector and increase the minimum value of $\chi^2$. However, then the bias in the flow parameters should be present in all direct sampling and pickup ion experiments as well as in remote-sensing studies, for example, observations of the heliospheric helium glow in 58.4~nm \citep[e.g.,][]{vallerga_etal:04a}.  

Yet another potential cause of the high $\chi^2$ value in the fits may be data points which could potentially be affected by intermittent systematic effects, as suggested by \citet{schwadron_etal:15a}. These authors fitted the positions of the observed peak of the ISN He signal using a model similar to ours. An important difference between the data treatment by \citet{schwadron_etal:15a} and this study is that while we fit the data averaged over the good time intervals for entire orbits, \citet{schwadron_etal:15a} (and, in fact, \citet{leonard_etal:15a} and \citet{mobius_etal:15b}) retain their time resolution and fit the time evolution of the latitude of the ISN~He beam. Based on this, they are able to identify the intervals of time when the peak positions differ from their fitted model during the portions of individual orbits. Based on this comparison, \citet{schwadron_etal:15a} point out a few short intervals when the observed peak positions systematically differ from their best-fit model. This may be indicative of an additional intermittent signal that is strong enough to shift the fitted peak position of the signal during a time interval of the order of hours. In the absence of this systematic effect, our minimum $\chi^2$ would be very unlikely to exceed the expected value due to purely statistical fluctuations because the Poisson scatter of the ISN He signal is taken into account in our uncertainty system. Detection of the hypothetical intermittent signal is very challenging in the data product we use since we have a signal integrated over entire good time intervals. The data were selected based on the well-justified criteria described earlier in the paper. We are reluctant to reject data based on their poor fit to the model. Instead, we rely on objective data selection criteria and we acknowledge that the filtering thus performed may not be perfect or that the adopted model may be too simple. 

The uncertainty scaling due to the excessively high minimum $\chi^2$ value should be considered in a situation such as our according to the literature \citep{rosenfeld:75a,press_etal:07a, olive_etal:14a}. The procedure assumes that the errors and uncertainties responsible for the excessively high minimum $\chi^2$ value are independent and Gaussian. This need not be the case, but, on the other hand, scaling the uncertainty to larger values using a well-defined scheme seems to be a better solution than neglecting the extra unknown uncertainty altogether. In our case, we have identified the main suspected cause of the overly high minimum $\chi^2$ value, i.e., the non-perfect model of the Warm Breeze, and we believe that before a more extensive analysis of this component is available, tentative adoption of the uncertainty assessment we propose in this article is the most reasonable solution.

\subsection{Comparison with other estimates}
Given the above insights, we believe that, for now, the data selection we have made is close to optimum. On one hand, it eliminates the orbits and spin-angle bins where the influence of the Warm Breeze and ISN~H is the largest while, on the other hand, it retains data points from all of the seasons in the sample, in approximately equal numbers on both sides of the ISN~He peak, and provides a relatively large total number of data points. Thus, we have a sufficient statistics and we do not spatially bias the data. As the most likely parameter set, we recommend the set obtained from fitting the full set using the direct-fitting method, with the uncertainties calculated by scaling the formal covariance matrices of the data by the $S$ values defined in Equation~\ref{eq:defRedChi}. 

The results from fitting the data from the first six {\it IBEX} ISN~He seasons are in a very good agreement with the results obtained by \citet{bzowski_etal:14a} and \citet{wood_etal:15a}, who analyzed all three seasons of {\it Ulysses/GAS} data. This agreement is both for the inflow velocity vector and for the temperature. Compared with the previous determination of ISN~He parameters by \citet{witte_etal:04a}, the velocity vector is supported within uncertainties, but the temperature is higher by more than 1000~K. Our present result is also in good agreement with the preliminary analysis of the 2013 and 2014 seasons by \citet{mccomas_etal:15a}, as well as with the present analysis by \citet{leonard_etal:15a}, \citet{mobius_etal:15b}, and \citet{schwadron_etal:15a}. Particularly good agreement among all of those studies was obtained for the velocity vector. The temperature obtained by \citet{mobius_etal:15b} is a little higher than in this study, but not outside the uncertainty range ehich they obtained. Moreover, similar to our findings,  \citet{schwadron_etal:15a} also find that the longitudes found for individual seasons vary, with the other ISN~He parameters sliding along the 4D correlation tube they adopted in their analysis. 

\citet{mobius_etal:15b} suggest that a good fraction of the scatter in the \citet{schwadron_etal:15a} results may be due to the statistical (Poisson) scatter of the data they used. The influence of Poisson noise on the analysis following the three-step method described in \citet{mobius_etal:15b} may be different than in our analysis. The data products used by \citet{mobius_etal:15b} and which we use differ in one important aspect. While \citet{mobius_etal:15b} utilize several subgroups distributed in time during all of the orbits, we take the counts integrated over entire ISN good time intervals, and so the counts per pixel that we have are much larger than in the data product used by \citet{mobius_etal:15a}. As pointed out by \citet{swaczyna_etal:15a}, in our case, the Poisson scatter is the dominant uncertainty in the low-count data points, which thus are weighed relatively little in the entire data system. For the higher count pixels, however, the Poisson uncertainties, equal to the square root of the total number of points in a given spin-angle bin, are very small compared with the total number of counts in the bins. To seriously bias the result we obtain, a group of points would need to fluctuate strongly (by percentage), which is very unlikely for the high total count. Perhaps because of this difference, the analysis method used by \citet{mobius_etal:15b} and \citet{schwadron_etal:15a} may be more sensitive to the Poisson scatter than ours. On the other hand, \citet{schwadron_etal:15a} found some contiguous series of statistically significantly enhanced count rates that do not fit to their model, and they suspect that these features may be due to an intermittent, statistically significant signal. This hypothetical source is not accounted for in our uncertainty system and, indeed, if present, may be one of the reasons for the relatively large scatter between the results from individual seasons and the large minimum $\chi^2$ value we obtained. 

\citet{redfield_linsky:08a} analyzed interstellar absorption lines in the spectra of nearby stars and concluded that the Sun is moving through the Local Interstellar Cloud at a relative velocity equal to $23.84 \pm 0.9$~\kms and the flow direction in  galactic coordinates is $l = 7.0\degr \pm 3.4\degr$, $b = 13.5\degr \pm 3.3\degr$. The temperature of $7500 \pm 1300$~K that they obtained is in excellent agreement with our present findings, as well with the {\it Ulysses} reanalysis by \citet{bzowski_etal:14a} and \citet{wood_etal:15a}, and the turbulent speed is $\xi = 1.62 \pm 0.75$~\kms. The velocity directions agree within the error bars (our velocity direction with the uncertainties scaled up using Equation~\ref{eq:defRedChi} is $ l = 3.77\degr \pm 0.22\degr$, $b = 14.95\degr \pm  0.41\degr$). The largest difference is in speed: ours is a larger by $\sim 2$~\kms, i.e., by more than the uncertainty in the \citet{redfield_linsky:08a} measurement. These differences may be due to local turbulence in the interstellar medium. The magnitude of the turbulent speed in the local interstellar medium of $\sim 2$~\kms is approximately in agreement with the values expected from \citet{redfield_linsky:04a}. If we this difference is indeed due to turbulence, then the {\it IBEX} and {\it Ulysses} measurements provide a lower limit for the spatial scale of this turbulence: it must be significantly larger than the path covered by the Sun during the Ulysses and {\it IBEX} measurements within the local interstellar matter, namely, $\sim 25\, {\mathrm{yr}} \times 5.3\,{\mathrm{ AU/yr}} = \sim 130$~AU. On the other hand, \citet{frisch_etal:02a} suggested that the Sun may be located in a boundary region between nearby clouds, in fact, in a region where a gradient in the local interstellar matter flow velocity persists. This gradient hypothesis was recently and independently reiterated by \citet{gry_jenkins:14a}. In this case, direct-sampling experiments, like GAS/Ulysses and {\it IBEX-Lo}, provide a unique anchor point for the system of flow velocities of the interstellar matter near the Sun, but are not directly comparable with the results of the sampling of this gas over parsec-scale lines of sight. 

\subsection{Thermal Mach number of the ISN~He flow as the most robust parameter obtained from {\it IBEX} observations}
The results we obtain from the global fit and from the fits for individual seasons are in excellent agreement with each other, with the results of our previous analyses of {\it IBEX} observations, and with the results of the recent analysis of {\it Ulysses} measurements \citep{bzowski_etal:14a, wood_etal:15a} in one interesting aspect. \citet{mccomas_etal:12b} found a correlation between the inflow longitude and the thermal Mach number of the ISN~He flow. For the inflow longitude of $259\degr$ which they used, the Mach number expected from this correlation is $5.013 \pm 6.5\%$, which differs by only $1.3\%$ from the value we obtain now based on a much larger data set and using a method without analytical simplifications. Also, the Mach number values resulting from the velocity and temperature reported by \citet{bzowski_etal:12a} agree well with our present result: the ISN~He Mach number of 4.98 agrees with the Mach numbers we obtain for the 2009 and 2010 seasons and is within $\sim 2.5\%$ of the Mach number from the present best-fit parameters.

Adoption of the ISN~He parameter values found by \citet{witte:04} in the original analysis of {\it Ulysses} observations ($v = 26.3$~\kms, $T = 6300$~K) results in a Mach number equal to 5.631, which differs from the present result by more than $10\%$, i.e., very significantly. However, the Mach number obtained from analysis of the {\it Ulysses} observations by \citet{bzowski_etal:14a} ($v = 26+1/-1.5$~\kms, $T = 7500+1500/-2000$~K) is equal to 5.103, and by \citet{wood_etal:15a} ($v = 26.08\pm0.21$~\kms, $T = 7260\pm270$~K) is 5.202. These values agree within $0.5\%$ and $2.4\%$, respectively, with those obtained in our present analysis. It seems then that the Mach numbers of the ISN~He inflow obtained from various analyses of direct-sampling observations by Ulysses and {\it IBEX} are consistent with the value resulting from our analysis, equal to $5.08 \pm 0.03$.  

\section{Summary and conclusions}
We have analyzed measurements from the first six ISN~He observation seasons of the {\it IBEX} using a sophisticated data uncertainty and parameter fitting system developed by \citet{swaczyna_etal:15a} and the newest version of the ISN~He simulation WTPM by \citet{sokol_etal:15b}. We identified a surprisingly strong influence of the Warm Breeze, recently discovered by \citet{kubiak_etal:14a}, and accordingly performed a careful data selection to obtain the cleanest sample of ISN~He observations. We used two alternative parameter fitting methods: (1) the direct flux fitting method, originally used by \citet{bzowski_etal:12a} to analyze {\it IBEX} measurements and by \citet{bzowski_etal:14a} to analyze {\it Ulysses} measurements of ISN~He, and currently strongly refined and optimized by \citet{swaczyna_etal:15a}, and (2) the Gaussian beam fitting method, which is an adaptation of the approach originally devised by \citet{lee_etal:12a} and \citet{mobius_etal:12a}. In addition to fitting the baseline data set, we also analyzed the data subset used by \citet{leonard_etal:15a} and \citet{mobius_etal:15b} and subsets from the individual observation seasons. 

We found that the ISN~He inflow velocity vector seems to be consistent with the velocity vector originally obtained by \citet{witte:04} from analysis of {\it Ulysses} measurements, but only for a temperature that is higher by at least 1100~K. We also found that the Mach number of the neutral He flow ahead of the heliosphere is equal to 5.08. These findings agree with the results of the {\it Ulysses} analysis by \citet{bzowski_etal:14a} and \citet{wood_etal:15a} with a very high accuracy of $2.5\%$, and with the recent study of {\it IBEX} results by \citet{mccomas_etal:15a}. The Mach number obtained now is in full agreement with the results obtained by \citet{bzowski_etal:12a}, \citet{mobius_etal:12a}, and \citet{mccomas_etal:12b}. 

The ISN~He speed and temperature are very highly correlated, with the correlation coefficient being equal to 0.95. Also, a very high correlation exists between the temperature and inflow longitude. This is because the parameters obtained from observations carried out in a plane close to the plane containing the inflow direction of ISN~He are located within a narrow tube in the 4D parameter space. Consequently, the uncertainties of the ISN~He parameters are much wider along the parameter correlation than across this line, as shown explicitly by \citet{mccomas_etal:12b} based on a simplified analytical theory by \citet{lee_etal:12a}. 

The ISN~He inflow parameters were obtained assuming that the ISN~He adheres to a spatially homogeneous Maxwell--Boltzmann distribution at a distance of 150~AU from the Sun. We refer to this parameter as the tracking distance and physically it can be regarded as the distance at which the test atom was placed into its final solar orbit due to a collision with the ambient member particles from the interstellar gas. This concept is introduced to the analysis to acknowledge the fact that the ISN~He gas is collisionless on spatial scales limited to $\sim150$~AU. The rationale for the adoption of a finite tracking distance and the influence of the tracking distance magnitude on the simulated ISN~He signal is presented by \citet{sokol_etal:15b}. The parameter search result is mildly sensitive to the adopted tracking distance, e.g., the speed may change by $\sim 0.3$~\kms, as shown by \citet{mccomas_etal:12b}, \citet{mobius_etal:15b} and \citet{mccomas_etal:15b}. 

We also found evidence that the data very likely contain an unaccounted component that manifests itself systematically in the residuals and the high value of minimum $\chi^2$ found in the fitting. We suspect that at least part of this component may be due to less than optimum knowledge and removal of the Warm Breeze. However, we do not rule out that the adopted physical model of the parent population for the observed atoms being a superposition of the homogeneous ISN~He and Warm Breeze Maxwell--Boltzmann functions may also be too simple. 

We obtained a relatively large spread in the ISN~He parameters derived from analysis of data from individual observation seasons, but statistical analysis showed that such a spread can be expected given the parameter uncertainties and correlations between them. Therefore, we believe that currently the best estimate of the ISN~He inflow velocity vector, temperature, and Mach number are those that we found from the analysis of all of the observation seasons together, and that a reliable uncertainty estimate is obtained from the covariance matrix of the fit, adjusted  for the excessively high value of minimum $\chi^2$ found from the fitting. Thus, remembering the possibile existence of an effect unaccounted for  in the model that could bias the result, the best estimate for the ISN~He temperature and inflow velocity vector from this work is $\lambda_{\mathrm{ISNHe}} = 255.8\degr \pm 0.5\degr$, $\beta_{\mathrm{ISNHe}} = 5.16\degr \pm 0.10\degr$, $T_{\mathrm{ISNHe}} = 7440 \pm 260$~K, $v_{\mathrm{ISNHe}} = 25.8 \pm 0.4$~\kms, and $M_{\mathrm{ISNHe}} = 5.08 \pm 0.03$,  with a very strong correlation between the parameters that follow a 4D tube in the parameter space. The estimates for the ISN~He Mach number show a very low relative uncertainty, and thus we consider them to be the most robust result of our analysis.

\acknowledgments
The portion of the research done in SRC PAS was supported by Polish National Science Center grant 2012-06-M-ST9-00455. A.G. and P.W. thank the Swiss National Science foundation for financial support. Work by the US authors was supported by the {\it IBEX} mission, which is part of NASA's Explorer Program.

\bibliographystyle{apj}
\bibliography{iplbib}{}

\begin{thebibliography}{}
\expandafter\ifx\csname natexlab\endcsname\relax\def\natexlab#1{#1}\fi

\bibitem[{{Banaszkiewicz} {et~al.}(1996){Banaszkiewicz}, {Witte}, \&
  {Rosenbauer}}]{banaszkiewicz_etal:96a}
{Banaszkiewicz}, M., {Witte}, M., \& {Rosenbauer}, H. 1996, \aaps, 120, 587

\bibitem[{{Bochsler} {et~al.}(2014){Bochsler}, {Kucharek}, {M{\"o}bius},
  {Bzowski}, {Sok{\'o}{\l}}, {Didkovsky}, \& {Wieman}}]{bochsler_etal:14a}
{Bochsler}, P., {Kucharek}, H., {M{\"o}bius}, E., {et~al.} 2014, \apjs, 210, 12

\bibitem[{{Bzowski} {et~al.}(2014){Bzowski}, {Kubiak}, {H{\l}ond},
  {Sok{\'o}{\l}}, {Banaszkiewicz}, \& {Witte}}]{bzowski_etal:14a}
{Bzowski}, M., {Kubiak}, M.~A., {H{\l}ond}, M., {et~al.} 2014, \aap, 569, A8

\bibitem[{Bzowski {et~al.}(2013)Bzowski, Sok{\'o}{\l}, Kubiak, \&
  Kucharek}]{bzowski_etal:13b}
Bzowski, M., Sok{\'o}{\l}, J.~M., Kubiak, M.~A., \& Kucharek, H. 2013, \aap,
  557, A50

\bibitem[{Bzowski {et~al.}(2012)Bzowski, Kubiak, M{\"o}bius, Bochsler, Leonard,
  Heirtzler, Kucharek, Sok{\'{o}}{\l}, H{\l}ond, Crew, Schwadron, Fuselier, \&
  McComas}]{bzowski_etal:12a}
Bzowski, M., Kubiak, M.~A., M{\"o}bius, E., {et~al.} 2012, \apjs, 198, 12

\bibitem[{{Frisch} {et~al.}(2002){Frisch}, {Grodnicki}, \&
  {Welty}}]{frisch_etal:02a}
{Frisch}, P.~C., {Grodnicki}, L., \& {Welty}, D.~E. 2002, \apj, 574, 834

\bibitem[{Frisch {et~al.}(2013)Frisch, Bzowski, Livadiotis, McComas,
  M{\"o}bius, Mueller, Pryor, Schwadron, Sok{\'o\l}, Vallerga, \&
  Ajello}]{frisch_etal:13a}
Frisch, P.~C., Bzowski, M., Livadiotis, G., {et~al.} 2013, Science, 341, 1080

\bibitem[{Frisch {et~al.}(2015)Frisch, Bzowski, Drews, Leonard, Livadiotis,
  McComas, M{\"o}bius, Schwadron, \& Sok{\'o}{\l}}]{frisch_etal:15a}
Frisch, P.~C., Bzowski, M., Drews, C., {et~al.} 2015, \apj, 781:61,
  10.1088/0004

\bibitem[{{Fuselier} {et~al.}(2014){Fuselier}, {Allegrini}, {Bzowski}, {Dayeh},
  {Desai}, {Funsten}, {Galli}, {Heirtzler}, {Janzen}, {Kubiak}, {Kucharek},
  {Lewis}, {Livadiotis}, {McComas}, {M{\"o}bius}, {Petrinec}, {Quinn},
  {Schwadron}, {Sok{\'o}{\l}}, {Trattner}, {Wood}, \&
  {Wurz}}]{fuselier_etal:14a}
{Fuselier}, S.~A., {Allegrini}, F., {Bzowski}, M., {et~al.} 2014, \apj, 784, 89

\bibitem[{Galli {et~al.}(2014)Galli, Wurz, Fuselier, McComas, Bzowski,
  Sok{\'o}{\l}, Kubiak, \& M{\"o}bius}]{galli_etal:14a}
Galli, A., Wurz, P., Fuselier, S., {et~al.} 2014, \apj, 796, 9

\bibitem[{{Galli} {et~al.}(2015){Galli}, {Wurz}, {Park}, {Kucharek},
  {M{\"o}bius}, {Schwadron}, {Sok\'{o}{\l}}, {Bzowski}, {Kubiak}, {Swaczyna},
  {Fuselier}, \& {McComas}}]{galli_etal:15a}
{Galli}, A., {Wurz}, P., {Park}, J., {et~al.} 2015, \apjs, 220:30,
  doi:10.1088/0067-0049/220/2/30

\bibitem[{{Gry} \& {Jenkins}(2014)}]{gry_jenkins:14a}
{Gry}, C., \& {Jenkins}, E.~B. 2014, \aap, 567, A58

\bibitem[{{Grzedzielski} {et~al.}(2010){Grzedzielski}, {Bzowski}, {Czechowski},
  {Funsten}, {McComas}, \& {Schwadron}}]{grzedzielski_etal:10b}
{Grzedzielski}, S., {Bzowski}, M., {Czechowski}, A., {et~al.} 2010, \apjl, 715,
  L84

\bibitem[{H{\l}ond {et~al.}(2012)H{\l}ond, Bzowski, M{\"o}bius, Kucharek,
  Heirtzler, Schwadron, O'Neill, Clark, Crew, Fuselier, \&
  McComas}]{hlond_etal:12a}
H{\l}ond, M., Bzowski, M., M{\"o}bius, E., {et~al.} 2012, \apjs, 198, 9

\bibitem[{Kubiak {et~al.}(2013)Kubiak, Bzowski, Sok{\'o\l}, M{\"o}bius,
  Rodr{\'i}guez, Wurz, \& McComas}]{kubiak_etal:13a}
Kubiak, M.~A., Bzowski, M., Sok{\'o\l}, J.~M., {et~al.} 2013, \aap, 556, A39

\bibitem[{{Kubiak} {et~al.}(2014){Kubiak}, {Bzowski}, {Sok{\'o}{\l}},
  {Swaczyna}, {Grzedzielski}, {Alexashov}, {Izmodenov}, {Moebius}, {Leonard},
  {Fuselier}, {Wurz}, \& {McComas}}]{kubiak_etal:14a}
{Kubiak}, M.~A., {Bzowski}, M., {Sok{\'o}{\l}}, J.~M., {et~al.} 2014, \apjs,
  213, 29

\bibitem[{Lallement \& Bertaux(2014)}]{lallement_bertaux:14a}
Lallement, R., \& Bertaux, J.~L. 2014, \aap, 565, A41

\bibitem[{Lee {et~al.}(2012)Lee, Kucharek, M{\"o}bius, Wu, Bzowski, \&
  McComas}]{lee_etal:12a}
Lee, M.~A., Kucharek, H., M{\"o}bius, E., {et~al.} 2012, \apjs, 198, 10

\bibitem[{Leonard {et~al.}(2015)Leonard, M{\"o}bius, Bzowski, Fuselier,
  Heirtzler, Kubiak, Kucharek, Lee, McComas, Schwadron, \&
  Wurz}]{leonard_etal:15a}
Leonard, T.~W., M{\"o}bius, E., Bzowski, M., {et~al.} 2015, \apj, 804:42,
  doi:10.1088/0004-637X/804/1/42

\bibitem[{{McComas} {et~al.}(2015{\natexlab{a}}){McComas}, {Bzowski}, {Frisch},
  {Galli}, {Izmodenov}, {Katushkina}, {Kubiak}, {Lee}, {Leonard}, {M{\"o}bius},
  {Park}, {Schwadron}, J.M., {Swaczyna}, {Wood}, \& {Wurz}}]{mccomas_etal:15b}
{McComas}, D., {Bzowski}, M.~{Fuselier}, S., {Frisch}, P., {et~al.}
  2015{\natexlab{a}}, \apjs, 220:22, doi:10.1088/0067-0049/220/2/22

\bibitem[{{McComas} {et~al.}(2015{\natexlab{b}}){McComas}, {Bzowski}, {Frisch},
  {Fuselier}, {Kubiak}, {Kucharek}, {Leonard}, {M{\"o}bius}, {Schwadron}, J.M.,
  {Swaczyna}, \& {Witte}}]{mccomas_etal:15a}
{McComas}, D., {Bzowski}, M., {Frisch}, P., {et~al.} 2015{\natexlab{b}}, \apj,
  801, 28

\bibitem[{{McComas} {et~al.}(2009{\natexlab{a}}){McComas}, {Allegrini},
  {Bochsler}, {Bzowski}, {Christian}, {Crew}, {DeMajistre}, {Fahr}, {Fichtner},
  {Frisch}, {Funsten}, {Fuselier}, {Gloeckler}, {Gruntman}, {Heerikhuisen},
  {Izmodenov}, {Janzen}, {Knappenberger}, {Krimigis}, {Kucharek}, {Lee},
  {Livadiotis}, {Livi}, {MacDowall}, {Mitchell}, {M{\"o}bius}, {Moore},
  {Pogorelov}, {Reisenfeld}, {Roelof}, {Saul}, {Schwadron}, {Valek},
  {Vanderspek}, {Wurz}, \& {Zank}}]{mccomas_etal:09c}
{McComas}, D.~J., {Allegrini}, F., {Bochsler}, P., {et~al.} 2009{\natexlab{a}},
  Science, 326, 959

\bibitem[{{McComas} {et~al.}(2009{\natexlab{b}}){McComas}, {Allegrini},
  {Bochsler}, {Bzowski}, {Collier}, {Fahr}, {Fichtner}, {Frisch}, {Funsten},
  {Fuselier}, {Gloeckler}, {Gruntman}, {Izmodenov}, {Knappenberger}, {Lee},
  {Livi}, {Mitchell}, {M{\"o}bius}, {Moore}, {Pope}, {Reisenfeld}, {Roelof},
  {Scherrer}, {Schwadron}, {Tyler}, {Wieser}, {Witte}, {Wurz}, \&
  {Zank}}]{mccomas_etal:09a}
---. 2009{\natexlab{b}}, \ssr, 146, 11

\bibitem[{{McComas} {et~al.}(2011){McComas}, {Carrico}, {Hautamaki},
  {Intelisano}, {Lebois}, {Loucks}, {Policastri}, {Reno}, {Scherrer},
  {Schwadron}, {Tapley}, \& {Tyler}}]{mccomas_etal:11a}
{McComas}, D.~J., {Carrico}, J.~P., {Hautamaki}, B., {et~al.} 2011, Space
  Weather, 9, 11002

\bibitem[{{McComas} {et~al.}(2012){McComas}, {Alexashov}, {Bzowski}, {Fahr},
  {Heerikhuisen}, {Izmodenov}, {Lee}, {M{\"o}bius}, {Pogorelov}, {Schwadron},
  \& {Zank}}]{mccomas_etal:12b}
{McComas}, D.~J., {Alexashov}, D., {Bzowski}, M., {et~al.} 2012, Science, 336,
  1291

\bibitem[{{M{\"o}bius} {et~al.}(2004){M{\"o}bius}, {Bzowski}, {Chalov}, {Fahr},
  {Gloeckler}, {Izmodenov}, {Kallenbach}, {Lallement}, {McMullin}, {Noda},
  {Oka}, {Pauluhn}, {Raymond}, {Ruci{\'n}ski}, {Skoug}, {Terasawa}, {Thompson},
  {Vallerga}, {von Steiger}, \& {Witte}}]{mobius_etal:04a}
{M{\"o}bius}, E., {Bzowski}, M., {Chalov}, S., {et~al.} 2004, \aap, 426, 897

\bibitem[{{M{\"o}bius} {et~al.}(2009{\natexlab{a}}){M{\"o}bius}, {Kucharek},
  {Clark}, {O'Neill}, {Petersen}, {Bzowski}, {Saul}, {Wurz}, {Fuselier},
  {Izmodenov}, {McComas}, {M{\"u}ller}, \& {Alexashov}}]{mobius_etal:09a}
{M{\"o}bius}, E., {Kucharek}, H., {Clark}, G., {et~al.} 2009{\natexlab{a}},
  \ssr, 146, 149

\bibitem[{{M{\"o}bius} {et~al.}(2009{\natexlab{b}}){M{\"o}bius}, {Bochsler},
  {Bzowski}, {Crew}, {Funsten}, {Fuselier}, {Ghielmetti}, {Heirtzler},
  {Izmodenov}, {Kubiak}, {Kucharek}, {Lee}, {Leonard}, {McComas}, {Petersen},
  {Saul}, {Scheer}, {Schwadron}, {Witte}, \& {Wurz}}]{mobius_etal:09b}
{M{\"o}bius}, E., {Bochsler}, P., {Bzowski}, M., {et~al.} 2009{\natexlab{b}},
  Science, 326, 969

\bibitem[{M{\"o}bius {et~al.}(2012)M{\"o}bius, Bochsler, Heirtzler, Kucharek,
  Lee, Leonard, Petersen, Schwadron, Valocvin, Wu, Bzowski, Kubiak, Fuselier,
  Saul, Wurz, McComas, \& Crew}]{mobius_etal:12a}
M{\"o}bius, E., Bochsler, P., Heirtzler, D., {et~al.} 2012, \apjs, 198, 11

\bibitem[{{M{\"o}bius} {et~al.}(2015{\natexlab{a}}){M{\"o}bius}, {Bzowski},
  {Fuselier}, {Heirtzler}, {Kubiak}, {Kucharek}, {Lee}, {Leonard}, {McComas},
  {Schwadron}, {Sok\'{o}{\l}}, \& {Wurz}}]{mobius_etal:15b}
{M{\"o}bius}, E., {Bzowski}, M., {Fuselier}, S.~A., {et~al.}
  2015{\natexlab{a}}, \apjs, 220:24, doi:0.1088/0067-0049/220/2/24

\bibitem[{{M{\"o}bius} {et~al.}(2015{\natexlab{b}}){M{\"o}bius}, {Bzowski},
  {Fuselier}, {Heirtzler}, {Kubiak}, {Kucharek}, {Lee}, {Leonard}, {McComas},
  {Schwadron}, {Sok{\'o}{\l}}, \& {Wurz}}]{mobius_etal:15a}
---. 2015{\natexlab{b}}, Journal of Physics: Conference Series, 577, 012019

\bibitem[{Olive {et~al.}(2014)}]{olive_etal:14a}
Olive, K., {et~al.} 2014, Chin.Phys., C38, 090001

\bibitem[{Press {et~al.}(2007)Press, Teukolsky, Vetterling, \&
  Flannery}]{press_etal:07a}
Press, W.~H., Teukolsky, S.~A., Vetterling, W.~T., \& Flannery, B.~P. 2007,
  Chapter 10 (Cambridge University Press)

\bibitem[{{Redfield} \& {Linsky}(2004)}]{redfield_linsky:04a}
{Redfield}, S., \& {Linsky}, J.~L. 2004, \apj, 613, 1004

\bibitem[{{Redfield} \& {Linsky}(2008)}]{redfield_linsky:08a}
---. 2008, \apj, 673, 283

\bibitem[{Rosenfeld(1975)}]{rosenfeld:75a}
Rosenfeld, A.~H. 1975, Ann.Rev.Nucl.Sci., 25, 555

\bibitem[{Saul {et~al.}(2012)Saul, Wurz, M{\"o}bius, Bzowski, Fuselier, Crew,
  Rodriguez, Leonard, McComas, Schwadron, Bochsler, \& Scheer}]{saul_etal:12a}
Saul, L., Wurz, P., M{\"o}bius, E., {et~al.} 2012, \apjs, 198, 14

\bibitem[{Saul {et~al.}(2013)Saul, Bzowski, Fuselier, Kubiak, McComas,
  M{\"o}bius, Sok{'o}{\l}, Rodr{\'i}guez, Scheer, \& Wurz}]{saul_etal:13a}
Saul, L., Bzowski, M., Fuselier, S., {et~al.} 2013, \apj, 767, 130

\bibitem[{{Schwadron} {et~al.}(2015){Schwadron}, {M{\"o}bius}, {Leonard},
  {Fuselier}, {Bzowski}, {Frisch}, {Heirtzler}, {Kubiak}, {Kucharek}, {Lee},
  {McComas}, {Rahmanifard}, {Sok\'{o}{\l}}, \& {Swaczyna}}]{schwadron_etal:15a}
{Schwadron}, N., {M{\"o}bius}, E., {Leonard}, T., {et~al.} 2015, \apjs, 220:25,
  doi:0.1088/0067-0049/220/2/25

\bibitem[{{Schwadron} {et~al.}(2013){Schwadron}, {Moebius}, {Kucharek}, {Lee},
  {French}, {Saul}, {Wurz}, {Bzowski}, {Fuselier}, {Livadiotis}, {McComas},
  {Frisch}, {Gruntman}, \& {Mueller}}]{schwadron_etal:13a}
{Schwadron}, N.~A., {Moebius}, E., {Kucharek}, H., {et~al.} 2013, \apj, 775, 86

\bibitem[{{Sok{\'o}{\l}} \& {Bzowski}(2014)}]{sokol_bzowski:14a}
{Sok{\'o}{\l}}, J.~M., \& {Bzowski}, M. 2014, ArXiv e-prints, arXiv:1411.4826

\bibitem[{{Sok\'{o}{\l}} {et~al.}(2015{\natexlab{a}}){Sok\'{o}{\l}}, {Kubiak},
  {Bzowski}, \& {Swaczyna}}]{sokol_etal:15b}
{Sok\'{o}{\l}}, J.~M., {Kubiak}, M., {Bzowski}, M., \& {Swaczyna}, P.
  2015{\natexlab{a}}, \apjs, 220:27, arXiv:1510.04869

\bibitem[{{Sok\'{o}{\l}} {et~al.}(2015{\natexlab{b}}){Sok\'{o}{\l}}, {Bzowski},
  {Kubiak}, {Swaczyna}, {Galli}, {Wurz}, {M{\"o}bius}, {Kucharek}, {Fuselier},
  \& {McComas}}]{sokol_etal:15a}
{Sok\'{o}{\l}}, J.~M., {Bzowski}, M., {Kubiak}, M., {et~al.}
  2015{\natexlab{b}}, \apjs, 220:29, arXiv:1510.04874

\bibitem[{{Swaczyna} {et~al.}(2015){Swaczyna}, {Bzowski}, {Kubiak},
  {Sok\'{o}{\l}}, {M{\"o}bius}, {Leonard}, {Heirtzler}, {Kucharek},
  {Schwadron}, {Fuselier}, \& {McComas}}]{swaczyna_etal:15a}
{Swaczyna}, P., {Bzowski}, M., {Kubiak}, M., {et~al.} 2015, \apjs, 220:26,
  doi:10.1088/0067-0049/220/2/26

\bibitem[{{Vallerga} {et~al.}(2004){Vallerga}, {Lallement}, {Lemoine},
  {Dalaudier}, \& {McMullin}}]{vallerga_etal:04a}
{Vallerga}, J., {Lallement}, R., {Lemoine}, M., {Dalaudier}, F., \& {McMullin},
  D. 2004, \aap, 426, 855

\bibitem[{{Witte}(2004)}]{witte:04}
{Witte}, M. 2004, \aap, 426, 835

\bibitem[{Witte {et~al.}(2004)Witte, Banaszkiewicz, Rosenbauer, \&
  McMullin}]{witte_etal:04a}
Witte, M., Banaszkiewicz, M., Rosenbauer, H., \& McMullin, D. 2004, \asr, 34,
  61

\bibitem[{Witte {et~al.}(1996)Witte, Rosenbauer, Banaszkiewicz, \&
  Fahr}]{witte_etal:96}
Witte, M., Rosenbauer, H., Banaszkiewicz, M., \& Fahr, H.~J. 1996, \ssr, 78,
  289

\bibitem[{{Wood} {et~al.}(2015){Wood}, {M{\"u}ller}, \&
  {Witte}}]{wood_etal:15a}
{Wood}, B.~E., {M{\"u}ller}, H.-R., \& {Witte}, M. 2015, \apj, 801:62,
  arXiv:1501.02725

\bibitem[{Zank {et~al.}(2013)Zank, Heerikhuisen, Wood, Pogorelov, Zirnstein, \&
  McComas}]{zank_etal:13a}
Zank, G.~P., Heerikhuisen, J., Wood, B.~E., {et~al.} 2013, \apj, 763, 20

\end{thebibliography}

\end{document}